\documentclass[aps,prl,twocolumn,floatfix,tightenlines,showpacs,notitlepage]{revtex4-1}
\usepackage{amsfonts}
\usepackage{amssymb}
\usepackage{amsthm}
\usepackage{amsmath} 
\usepackage{color,soul}
 \usepackage[english]{babel}
\usepackage{mathtools}
\usepackage{physics}
\usepackage{verbatim}
\usepackage{import}
\usepackage{epsfig}
\usepackage[caption=false]{subfig} % use for side-by-side figures
\usepackage{graphicx} 
\usepackage{comment}
\usepackage[caption=false]{subfig} 
\usepackage[breaklinks,colorlinks = true,linkcolor = blue,urlcolor=blue,citecolor=blue]{hyperref}
\usepackage{scalerel}
 \usepackage{float}
 \usepackage{bm}
\usepackage[normalem]{ulem}
\usepackage{nicefrac}
\newcommand{\pkg}{\mathbb{P}_{\scaleto{\mathrm{k}_{\scaleto{\mathrm{G}}{4pt}}}{5.5pt}}}
\newcommand{\kg}{\mathrm{k}_{\scaleto{\mathrm{G}}{3.5pt}}}
\newcommand{\lamG}{\lambda_{\scaleto{\mathrm{G}}{3.5pt}}}
\newcommand{\kI}{k_{\scaleto{\mathrm{I}}{3.5pt}}}
\newcommand{\kb}{k_{\beta}}

\newcommand{\icts}{International Centre for  Theoretical Sciences, Tata Institute of Fundamental Research,  Bangalore 560089, India}
\begin{document}
\title{On the thermalization of the three-dimensional, incompressible, Galerkin-truncated Euler equation}
\author{Sugan Durai Murugan}
\email{sugan.murugan@icts.res.in}
\affiliation{\icts}
\author{Samriddhi Sankar Ray}
\email{samriddhisankarray@gmail.com}
\affiliation{\icts}
\begin{abstract}

	The long-time solutions of the Galerkin-truncated three-dimensional,
	incompressible Euler equation relax to an absolute equilibrium as a
	consequence of phase space and kinetic energy conservation in such a
	finite-dimensional system. These thermalized solutions are
	characterised by a Gibbs distribution of the velocity field and kinetic
	energy equipartition amongst its (finite) Fourier modes. We now show,
	through detailed numerical simulations, the triggers for the inevitable
	thermalization in physical space and how the problem is reducible to an
	effective one-dimensional problem making comparisons with the more
	studied Burgers equation feasible. We also discuss how our
	understanding of the mechanism of thermalization can be exploited to
	numerically obtain dissipative solutions of the Euler equations and  
	evidence for or against finite-time blow-up in computer simulations.

\end{abstract}

\maketitle

Inviscid equations of hydrodynamics which are constrained to have a finite
number of Fourier modes leads to thermalized flows which are distinctly
different from our more accustomed viscous fluids.  This is because Liouville's
theorem ensures that the  projection of the inviscid equations on a finite set
of Fourier modes leads to, at long times, an inevitable thermalized, absolute
equilibrium Gibbs
state~\cite{hopf_statistical_1952,lee_statistical_1952,kraichnan_inertial_1967,kraichnan_helical_1973}.
Consequently, this is accompanied by an equipartition of kinetic energy across
Fourier modes
$\vec{k}$~\cite{cichowlas_evolution_2005,cichowlas_effective_2005,krstulovic_two-fluid_2008}
quite unlike the celebrated Kolmogorov scaling $\sim k^{-5/3}$ associated with
turbulence in three dimensions (3D) or the $k^{-2}$ scaling of the entropy
solution in the one-dimensional (1D) Burgers problem ~\cite{hopf_partial_1950}. Therefore such
thermalized fluids are amenable to well-established theories of equilibrium
statistical physics while being intrinsically chaotic. Recently, such nonlinear
Hamiltonian systems have been used to settle questions in many-body statistical
physics of ergodicity and mixing~\cite{murugan_many-body_2021} as well as,
admittedly in 1D,  understanding vexing questions of complex-time
singularities~\cite{rampf_eye_2022}. 

From the more specific vantage point of turbulence and fluid dynamics, the relevance of
such systems is more subtle and less immediately obvious. This is particularly
so for 3D turbulence where several fundamental questions remain unanswered.
Hence, in the absence of the many theoretical tools available for studying the
1D Burgers equation~\cite{bec_burgers_2007}, it is tempting to exploit the
advantages of a 3D Galerkin-truncated incompressible Euler equation to make
sense of real turbulent flows. Of course, superficially, such equilibrium
solutions are in stark contrast to those obtained in (driven-dissipative)
turbulence or in numerical solutions of the viscous Navier-Stokes equation. And
yet the truncated equation retain the same nonlinear triadic structure as the
parent inviscid partial differential equations or indeed, in three dimensions,
the viscous Navier-Stokes equation which model turbulent flows. Thus in many
ways the 3D Galerkin-truncated incompressible Euler equation is a compelling
link between ideas of statistical physics for a Hamiltonian system with
conserved dynamics~\cite{rose_ha_fully_1978,orszag_analytical_1970} and those which
describe the behavior of out-of-equilibrium, driven-dissipative, viscous
turbulent
flows~\cite{frisch_turbulence_1995,kraichnan_structure_1959,kolmogorov_local_1991}.  In
the last  couple of decades or so, since the work of L'vov {\it et
al.}~\cite{l'vov_quasi-gaussian_2002} and subsequently Frisch {\it et
al.}~\cite{frisch_turbulence_2012}, the generalisation of the idea of Galerkin
truncation to fine-tune triadic interactions has lead to a narrowing of the gap
between equilibrium statistical physics and turbulence. This, in particular,
has been used most importantly in deepening of our understanding of central
questions in 3D turbulence such as
intermittency~\cite{lanotte_turbulence_2015,lanotte_vortex_2016,buzzicotti_intermittency_2016,buzzicotti_lagrangian_2016,
ray_non-intermittent_2018,tom_revisiting_2017,picardo_lagrangian_2020} and the issue of bottlenecks in the energy
spectrum~\cite{frisch_hyperviscosity_2008,frisch_real-space_2013,banerjee_transition_2014}.  Most
recently, the possibilities of small-scale thermalization in real
flows~\cite{bandak_dissipation-range_2022} provides further impetus to studying the interplay of
equilibrium statistical physics and turbulence often in dimensions which are
not necessarily integer~\cite{fournier_d-dimensional_1978, celani_turbulence_2010}.

\begin{figure*}
  \includegraphics[width=1.0\linewidth]{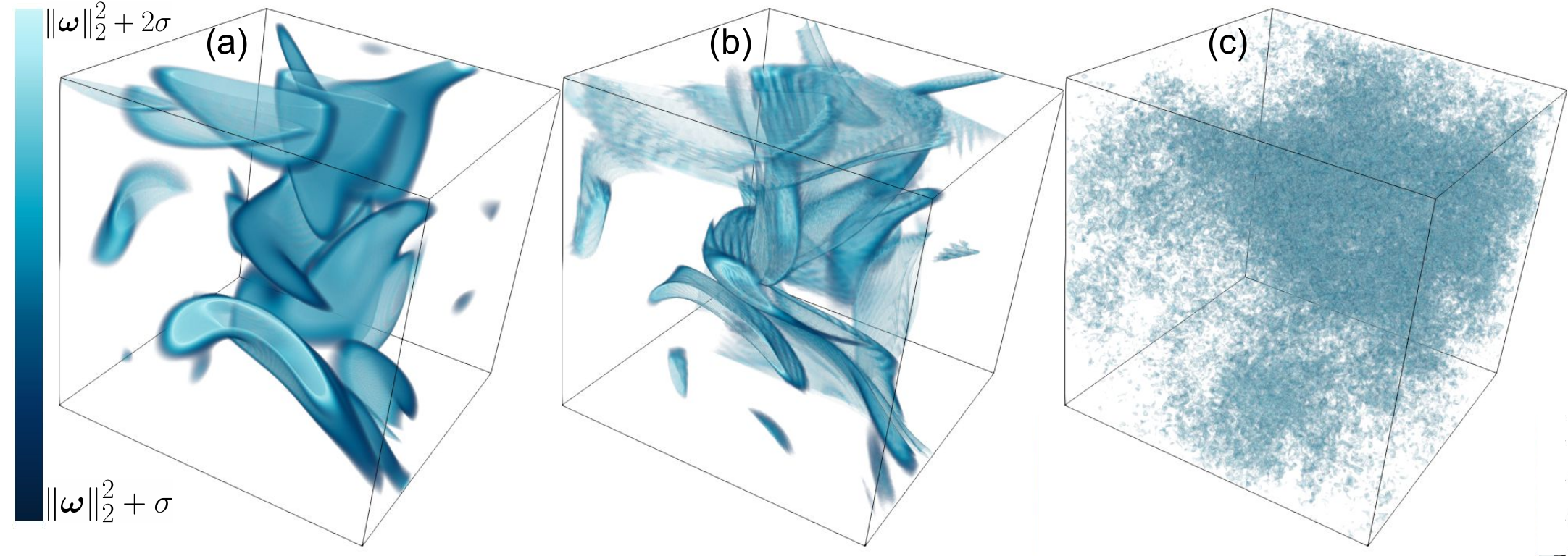}
	\caption{Isosurfaces of the vorticity field for $\sigma \leqslant  \abs{\bm{\omega}}^2 - \norm{ \bm{ \omega }} _2^2  \leqslant  2\sigma$  at (a) $t = 0.5$, 
	(b) $t = 0.85$ and (c) $t=2.5$. See \href{https://www.youtube.com/watch?v=pEOUlGQnrsQ&list=PL1TTBSO1jlfL49v8nDlpzoG9ZJg04mevA&index=6}{here} for an animation of the evolution of these isosurfaces from a non-thermalized to a fully thermalized state.}
  \label{fig:thermal}
\end{figure*}

There is another important reason why the Galerkin-truncated equation merits
attention.  One of the outstanding questions at the interface of physics and
mathematics is the existence of weak or dissipative
solutions~\cite{leray_motion_2016,onsager_statistical_1949} and the possibility
of finite-time blow-up for the 3D Euler
equation~\cite{gibbon_three-dimensional_2008,gibbon_three-dimensional_2008-1,frisch_singularities_2002}.
While a review of this subject goes well beyond the scope of the present paper,
suffice to say that probing the blow-up problem numerically is a
monumental challenge~\cite{brachet_small-scale_1983,brachet_numerical_1992,
boratav_reconnection_1992,kerr_evidence_1993,
shelley_dynamical_1993,boratav_direct_1994,kerr_velocity_2005,
cichowlas_evolution_2005,pelz_linearly_2005,hou_dynamic_2006,luo_potentially_2014,
moore_spontaneous_1979,morf_spontaneous_1980,
pelz_evidence_1997,gulak_high-symmetry_2005,chorin_evolution_1982,
siggia_collapse_1985,  pumir_collapsing_1990,  bell_vorticity_1992,
pelz_locally_1997, grauer_numerical_1991,grauer_adaptive_1998,
orlandi_nonlinear_2007,kolluru_insights_2022}.  Indeed, conjectures remain speculative at best
despite well-formulated
criterion~\cite{beale_remarks_1984,ponce_remarks_1985,constantin_geometric_1996,constantin_singular_2008,eyink_dissipative_2008,chae_finite-time_2007,deng_geometric_2005,deng_improved_2006}
which, in principle, should be easily detectable in well-resolved direct
numerical simulations (DNSs)~\cite{hou_blowup_2008}. The obstacle to this
however is that simulations are necessarily finite-dimensional:  The commonly
used spectral simulations~\cite{canuto_2012,leveque_1990} solve the
Galerkin-truncated and \textit{not} the infinite-dimensional partial
differential equations of inviscid flows. Hence, in finite times, which may
well precede the time of blow-up (as is for the inviscid one-dimensional
Burgers equation~\cite{ray_resonance_2011,venkataraman_onset_2017}), the
solutions thermalize (starting with the smallest scales) making methods for
singularity-detection, such as the analyticity strip approach~\cite{sulem_tracing_1983},
arduous~\cite{bustamante_interplay_2012,kolluru_insights_2022}.  Hence for finite
resolutions, in the absence of convergence of such truncated solutions (which
thermalize) to the actual (weak) solutions of the Euler equations themselves,
conjectures on blow-ups from
DNSs~\cite{beale_remarks_1984,constantin_geometric_1996,deng_improved_2006}
will remain unsettled till mechanisms to circumvent Gibbs states in
mathematically self-consistent ways are discovered.  The discovery of such
methods is of course contingent on knowing how truncated equations thermalize
in the first place. It is useful to recall that such methods have been
discovered for the more academic 1D Burgers
problem~\cite{pereira_wavelet_2013,farge_wavelet-based_2017,murugan_suppressing_2020,rampf_eye_2022,pereira_adaptive_2021}
owing to our thorough understanding of how the one-dimensional equation
thermalizes.

Thus the long-time chaotic, Gibbs solutions~\cite{murugan_many-body_2021} of the Galerkin-truncated Euler
equations play contrasting roles in studies of fundamental problems in
turbulence. On one hand, they allow us to connect ideas from statistical
physics to turbulence and on the other they remain a stumbling block in
numerical methods for studying questions of blow-up and dissipative solutions.
This makes the understanding of how such 3D flows thermalize particularly
essential.  As a result in recent years, since the pioneering work of Cichowlas
\textit{et al.}~\cite{cichowlas_evolution_2005}, a reasonably complete picture
of how energy equipartition happens in Fourier space has
emerged~\cite{cichowlas_effective_2005,krstulovic_two-fluid_2008,bos_dynamics_2006,krstulovic_cascades_2009}.
However, unlike the case of the 1D Burgers
equation~\cite{majda_remarkable_2000,ray_resonance_2011,pereira_wavelet_2013,venkataraman_onset_2017},
not much is known of the origins of thermalization in physical space for the 3D
problem. 

With this in mind, we perform detailed DNSs  of the unit
density Galerkin-truncated incompressible 3D Euler equation (Appendix A); the low-pass Galerkin projector $\pkg$ ensures that Fourier
modes of the velocity field are set to zero, via $\pkg {\bf u}({\bf x}) =
\displaystyle \sum_{|\bf{k}| \le \kg} e^{\imath {\bf k}\cdot{\bf x}}\hat{{\bf
u}}_{\bf k}$, for wavenumbers greater than a chosen threshold
Galerkin-truncation wavenumber $\kg$. 

While a long-time thermalized fluid, through Liuoville's theorem, with Gibbs statistics~\cite{murugan_many-body_2021} is obvious, the transition from a smooth initial 
condition which behaves like a ``viscous'' fluid for finite times to one which is 
thermalized and essentially devoid of structure, is far from obvious. A clue may be found in plots of the isosurfaces of the vorticity fields
as they evolve in time. In Fig.~\ref{fig:thermal}(a) we show a plot of the
vorticity $\qty(\bm{\omega} = \nabla \times {\bf u})$ isosurface for $\sigma \le \abs{\bm{\omega }}^2 - \norm{\bm{\omega}}_2^2 \le
2\sigma$, where $\sigma$ is the standard deviation of the enstrophy field, at early times
$\qty(t = 0.5)$ when the largest available wavenumbers are still not fully excited.  When
seen in the energy spectrum (Appendix A; Fig.~\ref{fig:spectra}) at the same time, there is no sign of thermalization. 
These enstrophy isosurfaces are smooth and indistinguishable from what
one would expect from an extremely high Reynolds number Navier-Stokes simulations
with similar initial conditions and at similar times.  At slightly later times, $\qty( t \gtrapprox 0.85)$
however isosurfaces show minute but detectable
oscillatory structures Fig.~\ref{fig:thermal}(b) with wavelengths $\lamG = 2\pi/\kg$, reminiscent
of what is seen for the corresponding problem in the one-dimensional Burgers
equation~\cite{ray_resonance_2011,venkataraman_onset_2017,clark_di_leoni_dynamics_2017}.
We recall that a similar phenomenon was seen recently in simulations of the 3D, Galerkin-truncated \textit{axisymmetric} 
  incompressible Euler equation~\cite{kolluru_insights_2022}.
These initially localised (in both physical and Fourier space)
oscillations rapidly spread through the domain, with increasing amplitudes,
whilst becoming non-monochromatic. A snapshot of these fully thermalized states
(Fig.~\ref{fig:thermal}(c)) looks noisy~\cite{pereira_adaptive_2021} with no resemblance to the well-formed
isosurfaces which characterise fully-developed turbulence or indeed solutions
of the truncated equation before the onset of thermalization
(Fig.~\ref{fig:thermal}(a)).  Consequently, the energy spectrum at such times
and beyond converges to a equipartition $E(k) \sim k^2$ (Appendix A; also \cite{cichowlas_evolution_2005}).

\begin{figure}
\includegraphics[width=1.0\linewidth]{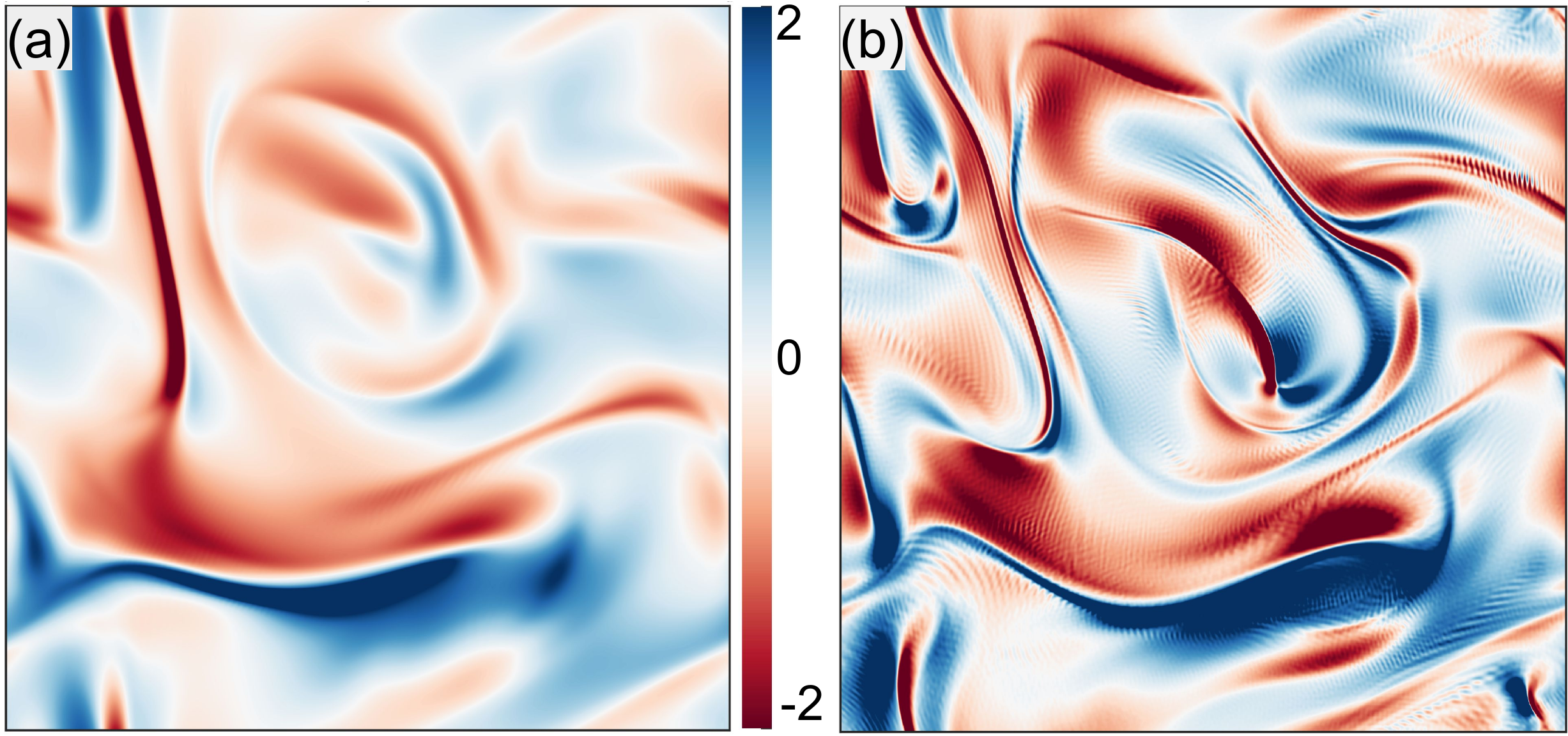}
\caption{Pseudo-color plots of the  strain field component ${\rm S}_{yz}$ in the $XY$ plane at times (a) before ($t = 1.2$) and (b)
after ($t = 1.8$) thermalization is triggered in the flow. 
While oscillatory structures are conspicuous by their absence for the former, 
coherent streaks of oscillations with wavelengths $\lamG$ are clearly visible for the latter. 
	See \href{https://youtu.be/4aql85cBYjE}{here} for an animation of the evolution of ${\rm S}_{yz}$ from a non-thermalized to a fully thermalized state.}
\label{fig:strain}
\end{figure}

While the signatures of thermalization are fairly obvious in plots such as
those shown in Fig.~\ref{fig:thermal}, the incipient thermalized phase is best
captured in visualisations of the velocity gradient. 
In Fig~\ref{fig:strain} we show two-dimensional ($XY$ plane)
cuts of the strain field ${\rm S}_{ij} \equiv
\partial u_i/\partial x_j$ which, at times when the effects of truncation are
felt, show clear, organised oscillatory structures (panel (b)) which are absent
at earlier times (panel (a)).
We recall that in the one-dimensional (1D) inviscid Galerkin-truncated Burgers
problem, the oscillatory structures that trigger thermalization are initially
localised at point(s) co-moving with the shock(s) through a resonance effect~\cite{ray_resonance_2011}. 
The flow we study now is fundamentally different: It is three-dimensional and
incompressible.  So how does thermalization kick in (Figs.~\ref{fig:thermal}(b) and ~\ref{fig:strain}(b))
the 3D Euler equations and is there an analogue of resonance points or do the
oscillations appear \textit{out of the blue}?

The answer to this is delicate and Fig.~\ref{fig:strain}(b) is suggestive.
Starting from initial conditions (such as the ones we have) which concentrates
energy at large scales, the nonlinearity of the systems generates smaller and
smaller scales in time and generates structures ranging from of vortex sheets
to tubes. As smaller and smaller scales get excited, many of these structures
can sharpen (as thin sheets or
tubes)~\cite{chorin_evolution_1982,chorin_turbulence_1986,bell_vorticity_1992}
with a characteristic length scale $\sim \kg^{-1}$. Such sharp structures,
analogous to shocks in the 1D Burgers equation, act as a source of
\textit{truncation waves} of wavelength $\lamG$---indeed the Fourier transform
of the projection operator is a wave with wavenumber $\kg$---which travel along
the directions in which such structures are compressed.  In the representative
snapshot shown in Fig.~\ref{fig:strain}(b) the oscillations appear with wave
vectors which, for this realization of the flow, are clearly normal to the 
intense structures seen in the domain. Of course, whether such
oscillations amplify or rapidly diminish in space and time is determined by the
nature of the strain field locally as we illustrate below.  For oscillations
which do survive, the nonlinearity allows other modes to get quickly excited
and the nonlocality of the incompressible equation allows a rapid spread of
these complex oscillations across the whole domain. This eventually leads to a
chaotic, thermalized fluid, bereft of structure, and equipartition of kinetic
energy across Fourier modes as illustrated in Fig.~\ref{fig:thermal}(c). 

This phenomenological picture, though compelling, is difficult to
\textit{prove} in numerical simulations with the generic initial conditions
that we use: The complexity of the spatial structures generated does not allow
an easy way to test the different ingredients which go into the argument
constructed above. In order to substantiate our theory, we resort to DNSs which
are controlled in a way to isolate the two different effects at play: The
sharpening of velocity gradients $\nabla u ~\sim 1/\kg$ and the consequent
onset of thermalization along specific direction(s) relative to such intense
structures.

\begin{figure*}
\includegraphics[width=1.0\linewidth]{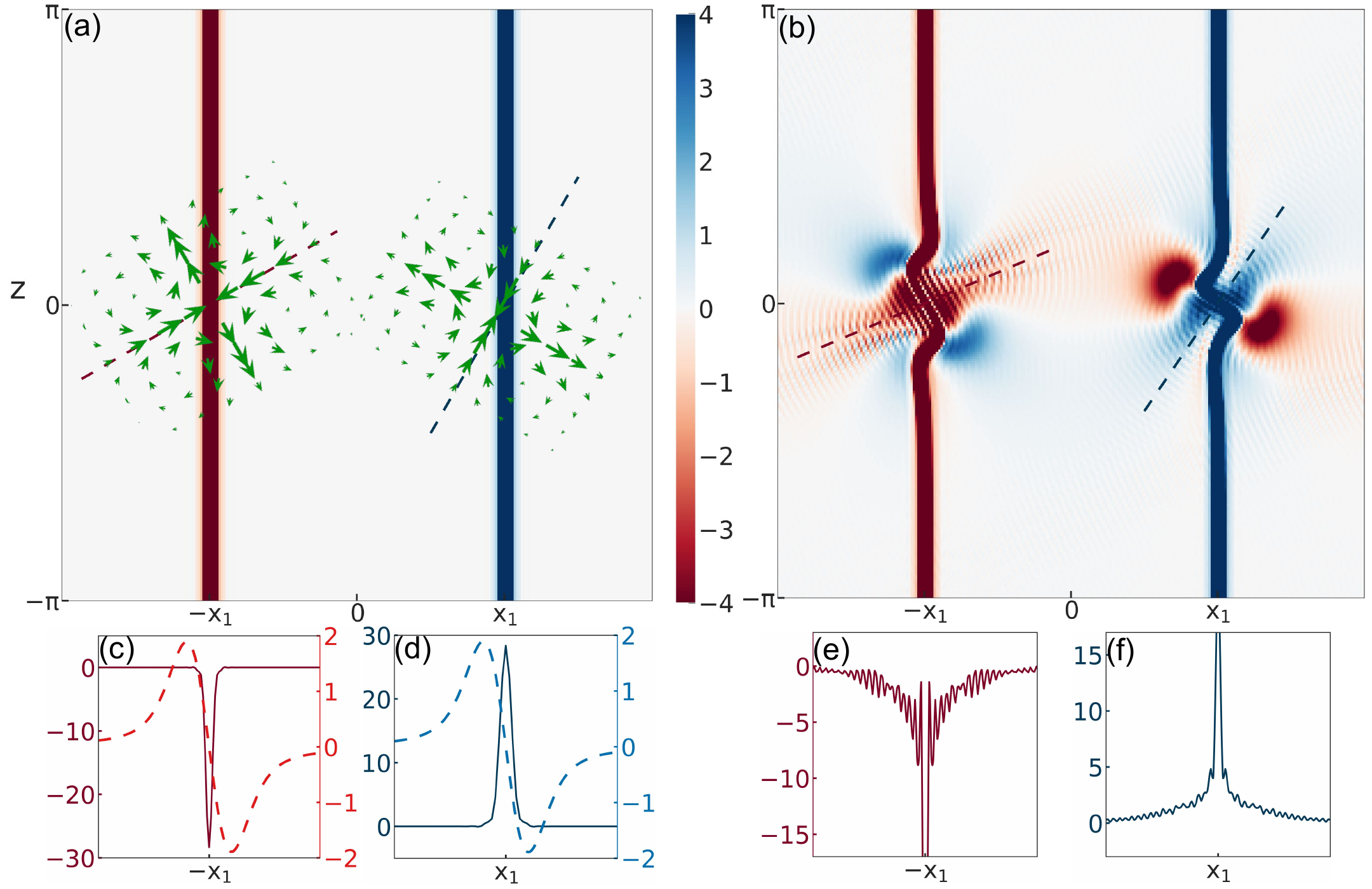}
	\caption{Pseudo-color plots of two-dimensional $XZ$ plane cuts of
	$\omega_z$ for the model flow (Appendix B; Eq.~\eqref{eq:IC1}) at (a)
	$t = 0$ and (b) $t = 0.15$ with their one-dimensional cuts (along dashed lines); the green 
	vectors in panel (a) corresponds to the velocity field.
	Along the compressional directions (in panel (a), shown as a dashed red and blue lines tilted at an angles $\pi /6$ and $\pi /3$ with respect to horizontal at $x=-x_1$ and $x_1$ respectively) the velocity component (dashed line) and $\omega_z$ (continuous line) 
	in panel (c) and (d) at $t=0$.
 	The corresponding plot for $\omega_z$ at $t = 0.15$ are shown in panels (e) and (f), respectively, showing 
	the oscillatory behaviour with wavelength $\sim \lamG$. See \href{https://www.youtube.com/watch?v=yY8oelqtIzA&list=PL1TTBSO1jlfL49v8nDlpzoG9ZJg04mevA&index=4}{here} 
	for an animation of the evolution of this flow to show the onset of thermalization.}
\label{fig:IC1}
\end{figure*}

Amongst the many candidate flows---such as isolated vortex tubes and sheets---we choose to work with 
an initial condition consisting of two separated, opposite-signed vortex sheets (in the YZ plane), located symmetrically 
at $x = x_1$ and $x = -x_1$, in a periodic box $\qty[-\pi ,\pi ]^3$. 
Furthermore, these sheets have a localised perturbing flow at their centers to disturb the sheet from equilibrium (Appendix B; Eq.~\eqref{eq:IC1}). 
This flow field, with the large scale background flow (which creates the sheet) suppressed (for clarity), is illustrated in Fig.~\ref{fig:IC1}(a) along with the two-dimensional 
velocity vectors which illustrate the regions of compression.

By using Fig.~\ref{fig:IC1}(a) as initial conditions, we solve the
Galerkin-truncated equation with $\kg = \nicefrac{N}{3}$.  Given the specific
configuration that we chose, the centers of both the sheets start to develop
sharper gradients: Along the compressional directions, at an angle $\pi/6$ at
$-{\bf x}_1$ and $\pi/3$ at ${\bf x}_1$ away from the horizontal (indicated in
Fig.~\ref{fig:IC1}(a) by blue and red dashed lines, respectively), the
perturbations lead to a squeezing of the vortex sheet.  As a result, these
structures sharpen and eventually the gradients become comparable to the
inverse of the truncation wavenumber. Thus, truncation waves are born as
explained before.

In the analogous, 1D Burgers problem these truncation waves emerge
from the region of the pre-shock and are constrained to travel along the
one-dimensional velocity field. Therefore, in the Burgers problem, it is
straightforward to identify the location of the oscillations: They ride on top
of the one-dimensional velocity field.  But for three-dimensional flows such as
ours, there are infinitely many possible directions along which these
oscillations emerge. Indeed, if such directions are chosen randomly by the
truncated dynamics, then the problem of thermalization and, crucially, finding
ways to circumvent becomes exceptionally hard. Fortunately, as we show below,
the solution to this is simpler and can be mapped to an effective
one-dimensional problem as we show below. 

Given these are three-dimensional flows, it is reasonable to conjecture that 
since the oscillations source from these sharp structures, they must be
constrained to be in the same direction along which the structure is
compressed. Thus the problem of knowing where in the flow the first signs of
thermalization appears is reduced to an effective one-dimensional problem along
very specific flow lines that generate sharp structures.  This conjecture is
easy to check for simpler flow geometries (such as in 
Fig.~\ref{fig:IC1}(a)). By design, it has a unique
direction along with the vortex sheets are compressed as clearly seen in the
streamlines in Fig.~\ref{fig:IC1}(a). This argument leads to the inevitable
conclusion that within a short time  oscillations of $\omega_z$,  with wave number $\kg$, should
appear along the compressional directions (Fig.~\ref{fig:IC1}(a)).

In order to test this hypothesis, we show, in Fig.~\ref{fig:IC1}(b) the
solution at time $t = 0.15$ of the truncated equation with the initial
conditions present in panel (a). We clearly see that, consistent with our
predictions, $\omega_z$ is oscillatory in the two directions of compression for
the two sheets. This hypothesis is further strengthened by performing
additional simulations where the sheets are compressed along arbitrary
directions (Fig.~\ref{fig:IC1-normal}). In all such cases,
consistent with our theory, $\omega_z$ is oscillatory in the same tilted
direction along which the compression happens. The corresponding problem for
the 1D Burgers equation is actually a special case of this phenomenon: In
one-dimensional space, the flow is compressional and hence the oscillations,
trivially seen in the velocity profile, accumulate at resonance points leading
to (at early times) spatially localised structures christened
\textit{tygers}~\cite{ray_resonance_2011}.

Therefore we now demonstrate, through numerical experiments with such
specialized initial conditions (see Appendix B for other examples), that the
onset of thermalization in the three-dimensional truncated system is
essentially one-dimensional: Monochromatic oscillations arise along the
compressional directions associated with fluid structures with critical
velocity gradients.  While this was implicit for the generic, large-scale
initial conditions which are used to solve the Galerkin-truncated Euler
equation, the use of such special flows is essential to make this phenomenon
evident. In more generic flows, such extreme velocity-gradient structures
proliferate the flow and emerge at different times. Hence each of these
structures act as independent sources of truncation waves and hence
thermalization.   

This observation of the precise mechanism at the heart of thermalization in 3D
flows is particularly important to devise numerical strategies to arrest
thermalization for the reasons discussed before. In Appendix C we suggest an
algorithm, based on first decomposing the ${\bm \omega}$ into local and
non-local contributions and subsequently reconstructing a vorticity field by
suppressing the former in a self-consistent way which preserves the small-scale
intense structures while discarding the oscillations. Our preliminary results (Fig.~\ref{fig:purge}(b)),
albeit based on such a \textit{static} filter (Appendix C) for the model flow, 
show encouraging signs that such approaches may well
diminish the precursor to small-scale thermalization and allow (a) dissipative
solutions and (b) extending the analyticity strip method for singularity
detection to longer times than currently possible. 

\begin{acknowledgements} We acknowledge insightful remarks by M. E. Brachet and R. Kerr and 
	important suggestions on the manuscript by R. Pandit and S. S. V. Kolluru. SSR is also indebted to U. Frisch 
	for several discussions on this subject over many years. The simulations were performed on the ICTS clusters
{\it Tetris}, and {\it Contra}. SSR acknowledges
	SERB-DST (India) projects MTR/2019/001553, STR/2021/000023 and CRG/2021/002766 for
financial support. SSR would like to thank the Isaac Newton Institute for Mathematical Sciences for support and hospitality 
	during the program \textit{Mathematical aspects of turbulence: where do we stand?} (EPSRC Grant Number EP/R014604/1) when part of 
	this work was done.  The authors acknowledges the support of the DAE, Govt. of
India, under project no.  12-R\&D-TFR-5.10-1100 and project no.  RTI4001
\end{acknowledgements}

\section{Appendix A: Direct Numerical Simulations and the Evolution of the Energy Equipartition Spectrum}

\newcommand{\hbAppendixPrefix}{A}
\renewcommand{\thefigure}{\hbAppendixPrefix\arabic{figure}}
\setcounter{figure}{0}
\renewcommand{\thetable}{\hbAppendixPrefix\arabic{table}} 
\setcounter{table}{0}
\renewcommand{\theequation}{\hbAppendixPrefix\arabic{equation}} 
\setcounter{equation}{0}

We perform DNSs of the unit-density, three-dimensional, Galerkin-truncated, incompressible ($\nabla \cdot {\bf u} = 0$) Euler equation
\begin{equation}
	\frac{\partial {\bf u}}{\partial t} = - \pkg{ \qty[ {\bf u}  \cdot \nabla {\bf u} - \nabla p ] }
	\label{eq:GT}
\end{equation}	
The low-pass Galerkin projector $\pkg$ sets to zero 
all modes of the velocity field with wavenumbers larger than 
the prescribed Galerkin-truncation wavenumber $\kg$:
$\pkg {\bf u}({\bf x}) =
\displaystyle \sum_{|\bf{k}| \le \kg} e^{\imath {\bf k}\cdot{\bf x}}\hat{{\bf
u}}_{\bf k}$. 

Our DNSs use a pseudo-spectral method with a fourth-order Runge-Kutta scheme for time
integration on $2\pi$ periodic domains with up to $N^3 = 512^3$ collocation
points and $\kg = \nicefrac{N}{3}$. Given the memory requirements for 3D visualisation of the 
$512^3$ domains, our plots in Fig.~\ref{fig:thermal} were from a $256^3$ simulation. However, the 2D slices 
which were used in Fig.~\ref{fig:strain} were taken from the larger $512^3$ data set. We have checked 
that our results and conclusions are consistent across simulations and choice of collocation points.
We choose initial conditions (also projected on the
compact Fourier domain) $E(k) \sim k^2 \exp(-\nicefrac{k^4}{\kI ^4})$ to ensure that the energy is concentrated
in the largest scales $\kI \sim \order{1}$. Galerkin truncation
ensures that the initial kinetic energy and phase space remains conserved
which, coupled with the finite-dimensionality imposed by the cut-off wavenumber
$\kg$, eventually leads to a thermalized fluid with kinetic energy
equipartitioned across all Fourier modes.

\begin{figure}
\includegraphics[width=1.0\linewidth]{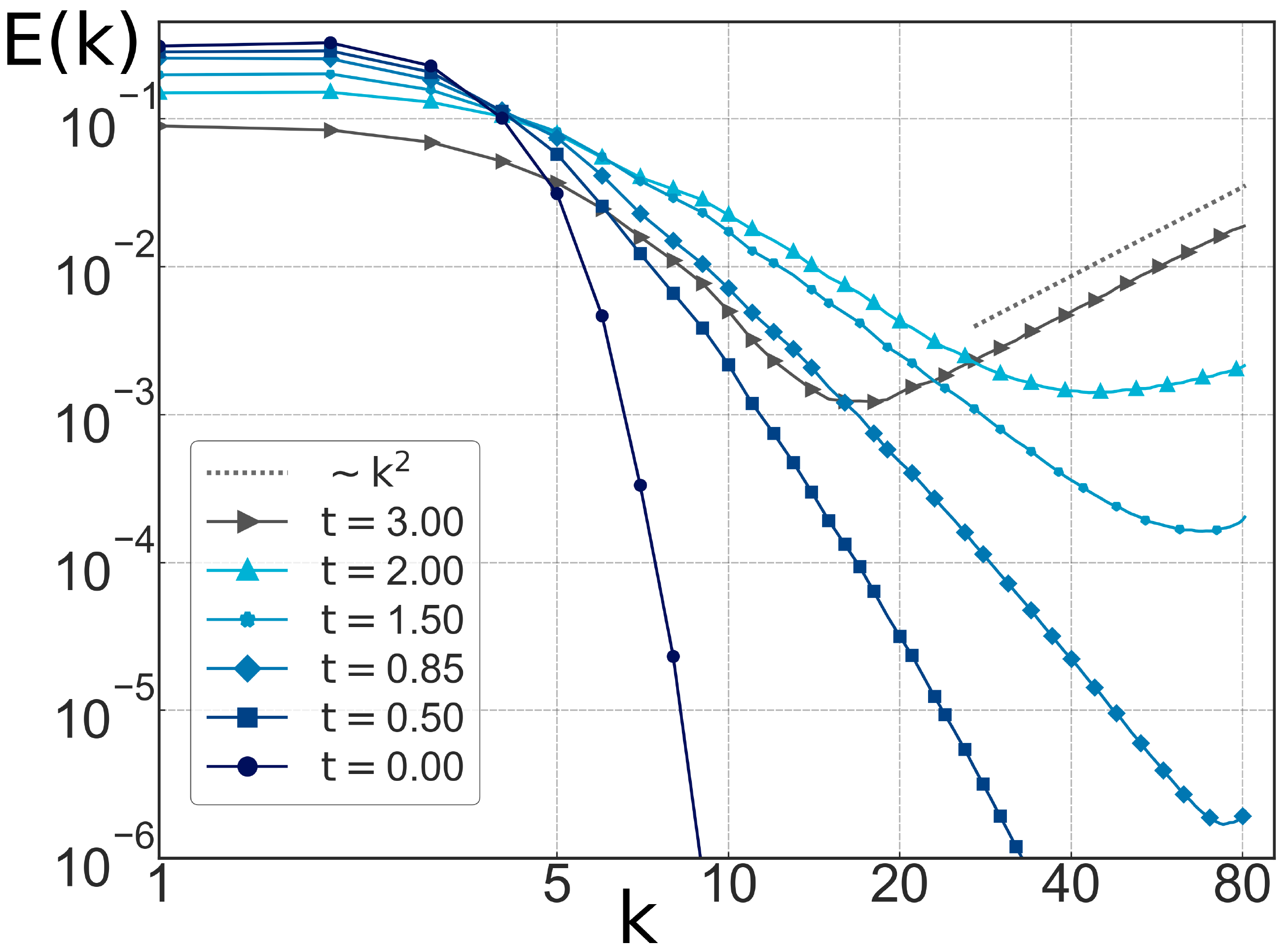}
	\caption{Loglog plots of the kinetic energy spectrum at different times from a DNS of the Galerkin-truncated Euler equation with generic, large-scale 
	initial conditions.}
\label{fig:spectra}
\end{figure}

Given the choice of initial conditions which confines kinetic energy at large scales, 
the excitement of the largest wavenumbers requires some time. In Fig.~\ref{fig:spectra} we 
show the evolution of the kinetic energy spectrum $E(k) \equiv \frac{1}{2}\sum_{k^\prime=k-1/2}^{k + 1/2} \left \langle {\bf u}({\bf k^\prime})\cdot {\bf u}({\bf k^\prime}) \right \rangle$ 
through representative loglog plots at various instances of time including those for which the vorticity isosurfaces 
have been shown in Fig.~\ref{fig:thermal}. Consistent with the emergence of oscillations when the largest wavenumbers are excited, 
we find the first signs of an equipartition spectrum $E(k) \sim k^2$ at times $t \gtrsim 0.85$. Similar evolution of the spectrum 
have been reported in the first study of this kind by Cichowlas \textit{et al.}~\cite{cichowlas_evolution_2005}.

\section{Appendix B: Simulations of Model Flows}

\renewcommand{\hbAppendixPrefix}{B}
\renewcommand{\thefigure}{\hbAppendixPrefix\arabic{figure}}
\setcounter{figure}{0}
\renewcommand{\thetable}{\hbAppendixPrefix\arabic{table}} 
\setcounter{table}{0}
\renewcommand{\theequation}{\hbAppendixPrefix\arabic{equation}} 
\setcounter{equation}{0}

\begin{figure*}
\includegraphics[width=1.0\linewidth]{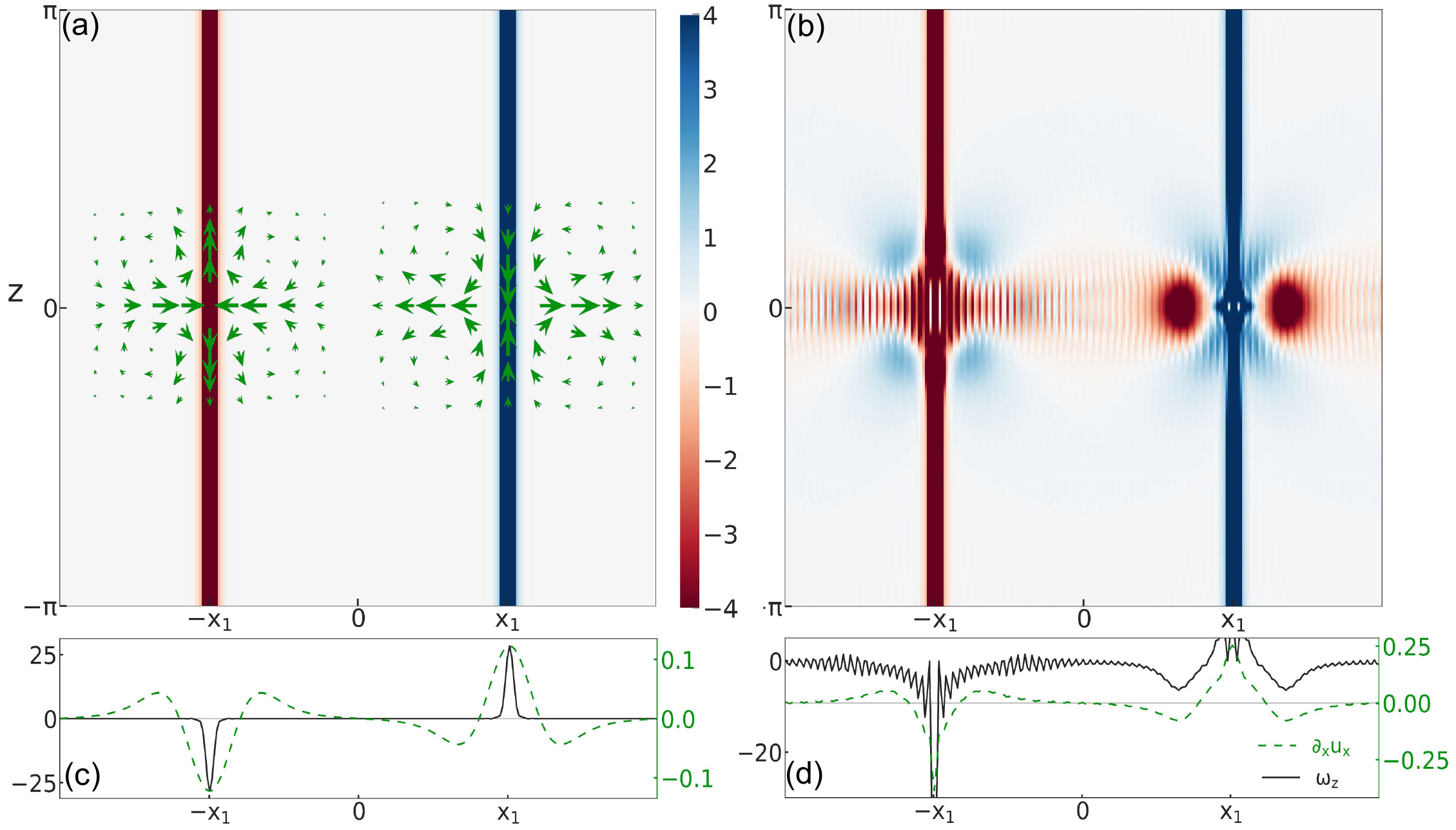}
	\caption{Pseudo-color plots of two-dimensional $XZ$ plane cuts of $\omega_z$ for the model flow (Eq.~\eqref{eq:IC1})  at 
	(a) $t = 0$ and (b) $t = 0.15$ with their one-dimensional cuts ($z = 0$) and those of the velocity gradient $\partial_x u_x$ at the 
	corresponding  times in panels (c) and (d), respectively. The initial profiles (panels (a) and (c)) are devoid of the $\lamG$ wavelength oscillations 
	which show up at later times along the directions of compression indicated by the 
	velocity vectors (green arrows) in panel (a). These oscillations are conspicuous, as clearly seen in panel (d) in regions of 
	of positive strain: $\partial_x u_x \ge 0$.}
\label{fig:IC1-normal}
\end{figure*}

In the main text, we report results from simulation~\ref{fig:IC1} of two separated vortex sheets with a localised perturbation. 
Such a flow configuration is generated through the following 
initial condition for $0\leqslant x \leqslant \pi $:
\begin{subequations}
\begin{align}
  u_x &= \mathcal{P}_{\perp}\qty[u_0 \kb (x-x_1) \exp(-\frac{1}{2}\kb ^2 \qty( \qty(x-x_1)^2 + y^2 + z^2 ) ) ] \\
  u_y &= \sqrt{2} \tanh\qty[\gamma k_G\qty(x-x_1)]\\
  u_z &= \mathcal{P}_{\perp}\qty[u_0 \kb z \exp(-\frac{1}{2} \kb^2 \qty( \qty(x-x_1)^2 + y^2 + z^2  ) )  ]. 
\end{align}
  \label{eq:IC1}
\end{subequations}
By symmetry, for $-\pi \leqslant x\leqslant 0 $ the velocity field is given by  $u_i(x,y,z)=\qty(1-2\delta_{iz})u_i(-x,y,z)$.
To ensure the incompressibility, the projection operator $\mathcal{P}_{\perp}\qty[\vb{f}]= \qty(1 - \qty(\nabla_{\perp} ^{-2})\grad_{\perp} \qty(\div_{\perp}{\vb{f}}))$ in the $XZ$ plane $\qty(\nabla _{\perp}=\qty{\partial _x,0,\partial _z})$ is applied to the $x$ and $z$ velocities.
The disturbance here is localized at $\vb{x}_1=\qty(x_1,0,0)$ and $\vb{-x_1}$; consequently the vortex sheet is stretched for the former and compressed for the latter.
The parameter $\gamma$ controls the intensity of the vortex sheet and is chosen to be 1/4 to suppress any inherent Gibbs oscillations which arise as $\gamma \to 1$. 
We fine-tune the extent of localisation of the perturbation through $\kb$ which, for the results presented here, was set to 4. Finally, the flow amplitude $u_0 = 5$ 
sets the energy of perturbation field ($\sim 10^{-3}$ relative to that of the vortex sheet) as well as the time-scale. 

\begin{figure}
\includegraphics[width=1.0\linewidth]{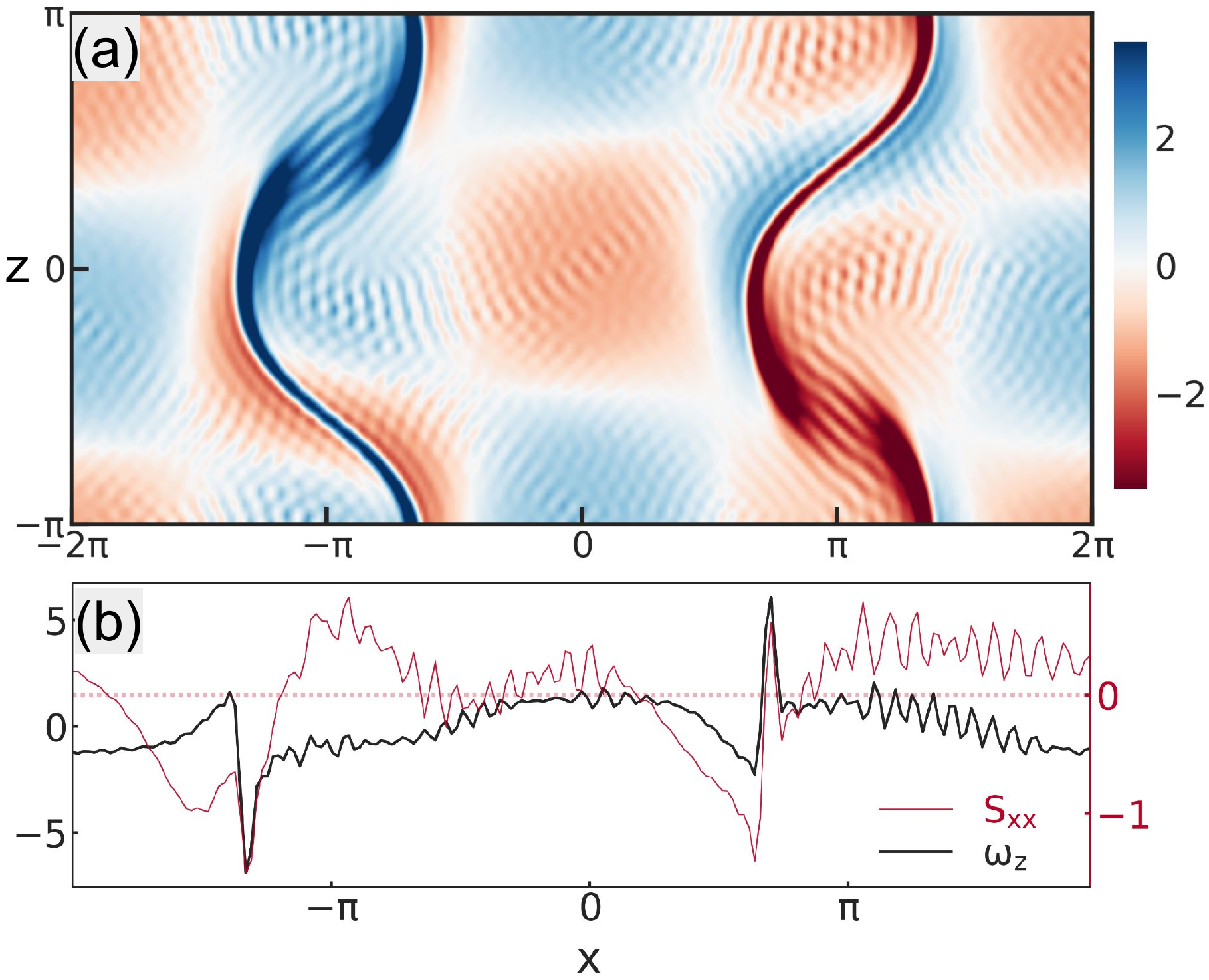}
	\caption{(a) A pseudo-color plot of the two-dimensional $XZ$ plane cut of
	$\omega_z$ for a pair of vortex sheets advected by a Taylor-Green 
	flow at time $t = 1.3$. (b) The one-dimensional cut of $\omega_z$ and 
	the strain component $S_{xx}$ along the thick, dashed horizontal line of panel (a) shows 
	the strong correlation between the thermalization-inducing oscillations and positive 
	values of $S_{xx}$.}
\label{fig:TG}
\end{figure}

In order to test the robustness of the claim and conclusions drawn from Fig.~\ref{fig:IC1}, we rotate the disturbance field 
in arbitrary directions to see if the oscillations in $\omega_z$ picks out these directions every time. In the main text 
we had shown (Fig.~\ref{fig:IC1}) results from an instance where 
the disturbance field $u_x,u_z$ is rotated by 
$\theta =\nicefrac{\pi }{6}$ from the normal of the sheet in the $XZ$ plane for the left half of the domain with its symmetric counterpart in 
the right half. 
To further underline this point in Fig.~\ref{fig:IC1-normal} we show results from simulations where $\theta = 0$; 
in agreement with the conjectures in the main text, we find that 
$\omega_z$ is clearly oscillatory along these new directions of compression which are normal to the 
vortex sheet at $-{\bf x}_1$. 

In the model flow discussed above and in the main text, the parallel vortex
sheets were subject to imposed perturbations. However, to make the system more
realistic, we now generalise this by immersing the two parallel vortex sheets 
(setting $u_x = u_z = 0$ in Eq.~\eqref{eq:IC1})
in a background Taylor-Green velocity field~\cite{cichowlas_effective_2005} (Fig.~\ref{fig:TG}(a)) and evolve
this system in time by using the Galerkin-truncated Euler equation. Unlike the
imposed localised perturbations before (Figs.~\ref{fig:IC1}
and~\ref{fig:IC1-normal}), in this case it is the evolution of a large-scale
background Taylor-Green flow that causes the sheets to \textit{bend} leading to
sharper gradients and eventually triggering off $\lamG$-wavelength oscillations
in the vorticity field in a manner similar to what we have already seen. In
Fig.~\ref{fig:TG}(a) we show a representative snapshot of this field (at time
$t = 1.5$). It is immediately obvious that for the more complicated form of
this flow, there are several different sources of truncation waves and
directions of compression leading to a greater proliferation of thermalization
\textit{hot spots} when compared to Figs.~\ref{fig:IC1}(a)
and~\ref{fig:IC1-normal}(b) which were curated especially to isolate single
directions of compression associated with each sheet. 

The use of the background Taylor-Green flow has one additional advantage. In
Fig.~\ref{fig:TG}(b) we show the vorticity component $\omega_z$ along the
middle of the domain ($z=0$ line in Fig.~\ref{fig:TG}(a)).
We immediately see the absence of oscillations to the left of the two sheets. This is
because in these regions the velocity gradient, along the direction of the cut $S_{xx}$,
is negative leading to the suppression of oscillations borne out of
compression.

\begin{figure}
\includegraphics[width=1.0\linewidth]{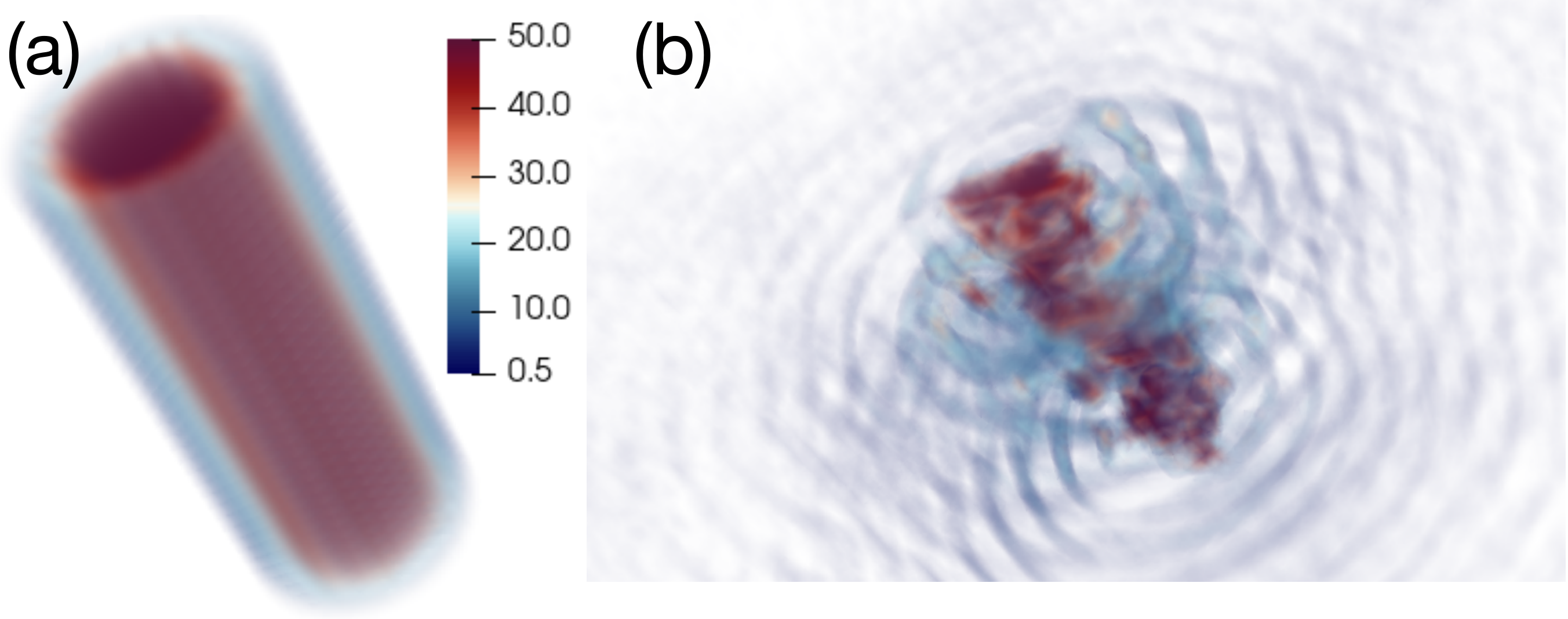}
	\caption{Pseudo-color plots of $\omega_z$ at (a) $t= 0$ and (b) $t = 0.8$; the axis of the vortex 
	filament is aligned along the $z$ direction.}	
\label{fig:Tube}
\end{figure}

As a final example, we simulate a vortex filament (Fig.~\ref{fig:Tube}(a)) in cylindrical 
coordinates: 
\begin{equation}
	\omega_z(r) \sim  \gamma \kg \exp\left[-(\gamma \kg r)^2\right ].
\label{eq:filament}
\end{equation}
Once again, this cylindrical vortex, whose thickness is determined by $\gamma = 0.25$, is immersed 
in a large-scale background flow which perturbs the filament. The precise form of this 
large-scale flow is irrelevant for the problem being studied.
Not surprising the evolution of this initial condition with the truncated Burgers 
equations leads to oscillations which are radial with the filament at its core~\ref{fig:Tube}(b).

\section{Appendix C: Suppressing Thermalization Through a Vorticity-field Decomposition}

\renewcommand{\hbAppendixPrefix}{C}
\renewcommand{\thefigure}{\hbAppendixPrefix\arabic{figure}}
\setcounter{figure}{0}
\renewcommand{\thetable}{\hbAppendixPrefix\arabic{table}} 
\setcounter{table}{0}
\renewcommand{\theequation}{\hbAppendixPrefix\arabic{equation}} 
\setcounter{equation}{0}

Understanding how finite-dimensional equations of hydrodynamics thermalize is
one aspect of this study; but perhaps the more important question relates to
whether this understanding can be exploited to devise more efficient algorithms
for numerical constructions of dissipative solutions of the Euler equations and
indeed conjectures for finite-time blow-up through methods such as the
analyticity strip~\cite{sulem_tracing_1983,bustamante_interplay_2012,kolluru_insights_2022}.

Operationally this would involving suppressing the oscillations which trigger
the flow to thermalize---making analyticity strip approaches to singularity
detection impractical~\cite{bustamante_interplay_2012}---and ensure
conservation of energy and thus the lack of dissipative solutions. From our DNSs 
it seems that a useful starting point would be a suitable filtering of the 
strain field (Fig.~\ref{fig:strain}) to remove the oscillatory structures.

To achieve this we adapt the method developed by Hamlington \textit{et al.}~\cite{Hamlington}
to decompose the strain field in to a local and non-local (background) contribution. 
This is trivially done for the vorticity field in Fourier space via 
\begin{align}
	\hat{\omega}^{\mathrm{(NL)}}(\vb{k})&= f(kR)\hat{\omega}(\vb{k})\\
  \hat{\omega}^{\mathrm{(L)}}(\vb{k})&= \hat{\omega} - \hat{\omega}^{\mathrm{(NL)}}
\label{strain_decomp}
\end{align}
where the \textit{hat} denotes Fourier space, the subscripts $L$ and $NL$ stand for the ``local'' and ``non-local'', respectively, $k = \abs{\vb{k}}$ and the filter
\begin{equation}
f(kR)= \dfrac{3\qty(\sin(kR)-kR\cos(kR))}{\qty(kR)^3} 
\end{equation}
is the Fourier transform of the three-dimensional complementary Heaviside function in spherical co-ordinates. Such a filter, by definition, ensures 
that the function on which it acts---namely the vorticity field in this case---is \textit{smoothened} by averaging out over a sphere of radius $R = \lamG$.
Evidently, the local contribution $\bm{\omega}^{\mathrm{L}}$ alone contains all the oscillations and hence the ``reconstructed'' field $\bm{\omega}^{\ast} \equiv \bm {\omega}^{\mathrm NL}$ 
with $\bm{\omega}^{\mathrm{L}}$ suppressed should be free of oscillations. Hence such a \textit{dynamic} filtering technique, namely, solving the truncated 3D Euler by recovering $\bm{\omega}^{\ast}$ 
and using this field to evolve at every time-step, should yield a non-thermalizing, dissipative flow. 

However, such an approach has the disadvantage that along with the oscillations, the small-scale, intense vortical structures are lost as well. We therefore adapt this idea of decomposing the 
field in a way which preserves the small-scale structures as far as possible and yet suppress the oscillatory triggers of thermalization. 
Thus, we propose:
\begin{align}
  \bm{\omega }^{\ast}(\vb{x})&= \bm{\omega }^{\mathrm{(NL)}}(\vb{x})+ \Gamma_{2m} (\vb{x})\bm{\omega }^{\mathrm{(L)}}(\vb{x})\\
  \Gamma_{2m} (\vb{x}):&= \mathrm{erf}\qty[\dfrac{\abs{\bm{\omega }}^{2m}}{\norm{\bm{\omega }}_{2m}^{2m}}]. 
  \label{vorticity_reconstructed}
\end{align}
The additional regularisation parameter $\Gamma_{2m}$  allows us to capture the essential, intense local vortical regions while still filtering out the oscillations in the flow. 
The $L_{2m}$ norm used in the definition of $\Gamma _{2m}$ further controls threshold level of that vortical regions we want to retain in the reconstructed field.

\begin{figure}
\includegraphics[width=1.0\linewidth]{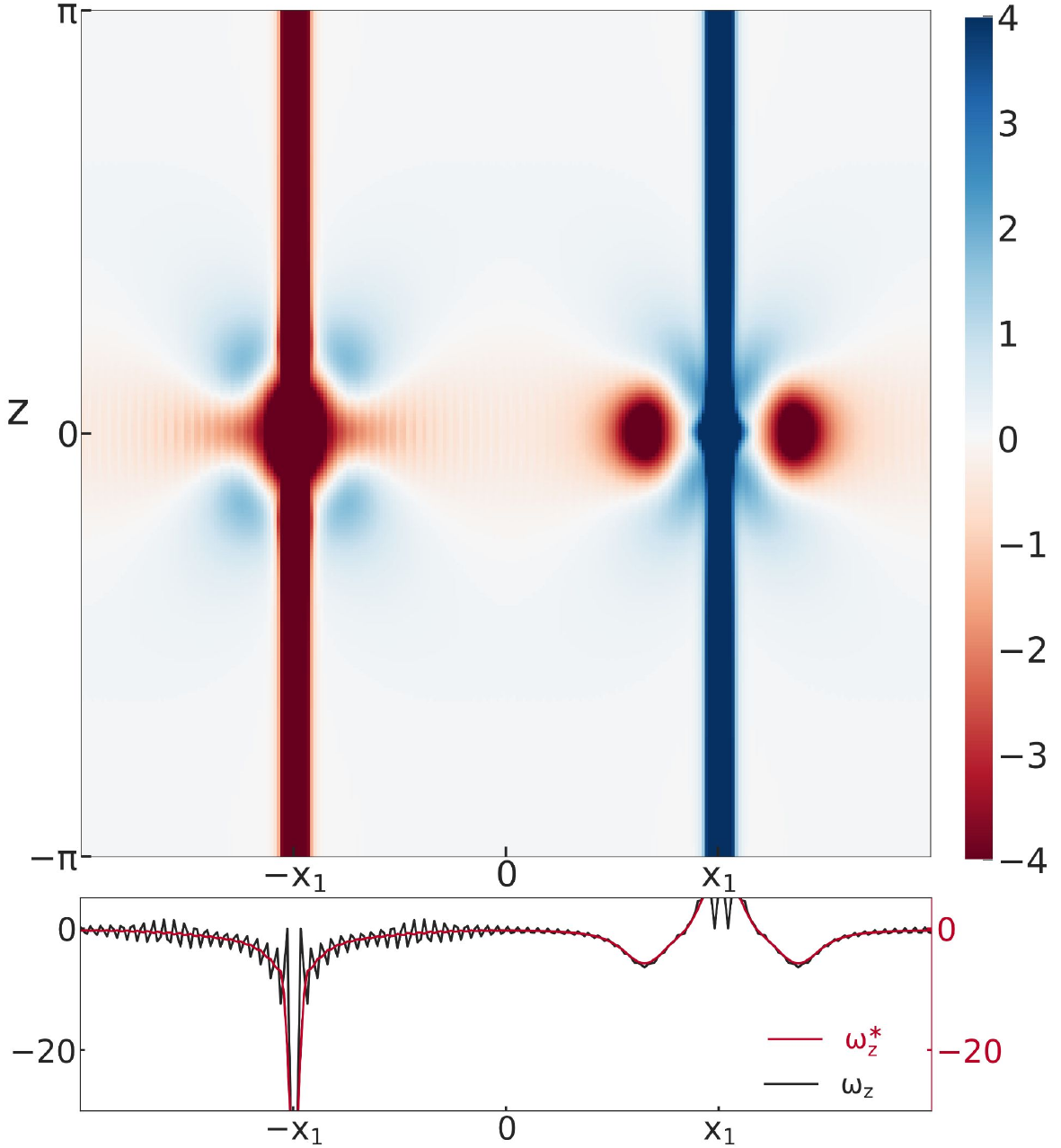}
	\caption{(a) Pseudo-color plot of the two-dimensional $XZ$ plane cut of the reconstructed vorticity field 
	$\omega^{\ast}_z$ for the model flow (Eq.~\eqref{eq:IC1}) at 
	$t = 0.15$. A comparison with the corresponding figure (Fig.~\ref{fig:IC1}(b)) of truncated simulation shows a 
	significant suppression of the oscillations. This is quantified in (b) plots of the one-dimensional cuts (along $z = 0$) $\omega_z$ (same as in 
	Fig.~\ref{fig:IC1}(d)) and $\omega^{\ast}_z$, as a function of $x$, along the center of the domain.} 
\label{fig:purge}
\end{figure}

While this method needs to be refined and rigorously examined in future studies for generic flow fields, we provide results from preliminary tests conducted on the model 
flow defined by Eq.~\eqref{eq:IC1}. 
In Fig.~\ref{fig:purge}(a) we show the reconstructed vorticity field at $t = 0.15$ for $m=4$, corresponding to the snapshot shown in Fig.~\ref{fig:IC1-normal}(b). 
A visual comparison of the two vorticity fields show that our reconstruction strategy indeed leads to a significant reduction in the oscillations while still preserving the intense structures, 
namely the vortical sheets in this case. This is quantified, in Fig.\ref{fig:purge}(b), by comparing the $z$-component of the vorticity along the $x$-axis ($z = 0$) in the middle of the domain for the truncated ($\omega_z$) and 
reconstructed fields ($\omega^{\ast}_z$). We clearly see that the oscillations responsible for thermalization, seen in $\omega_z$ more or less vanish on reconstruction as seen in the plot of ($\omega_z$). 
Furthermore our use of the regularisation parameter $\Gamma_{2m}$ does preserve fully the intense structure in the form of vortex sheets as seen by the near overlap of $\omega_z$ and $\omega^{\ast}_z$ at $-{\bf x}_1$ and 
${\bf x}_1$.

While Fig.\ref{fig:purge}(b) seems to underline the success of this
strategy---at least for such a curated flow---the illustrative flow field shown
in panel (a) still retains some traces of the oscillations. There are at least
two reasons why this is so. Firstly in our tests we have not filtered and
reconstructed the field at every time step but, as a proof of principle now,
used this as a \textit{static} filter and reconstruction at $t = 0.15$. A
dynamic filter, as discussed above, is essential and, perhaps, the
frequency---the time-intervals between successive filtering---with which the
filter should be applied needs further investigation. The latter may well be a
delicate point as shown in Ref.~\cite{murugan_suppressing_2020} for Fourier
space purging in the 1D Burgers equation.

Secondly, our preliminary explorations with different sharpness of the
regularisation parameter $\Gamma_{2m}$, shows that this, not surprisingly, is
critically important for more effective suppression of thermalization hotspots,
especially in the vicinity of flow structures with intense gradients. This will
become crucial when such strategies are investigated systematically in generic
3D flows.

\bibliography{references}

\begin{thebibliography}{84}%
\makeatletter
\providecommand \@ifxundefined [1]{%
 \@ifx{#1\undefined}
}%
\providecommand \@ifnum [1]{%
 \ifnum #1\expandafter \@firstoftwo
 \else \expandafter \@secondoftwo
 \fi
}%
\providecommand \@ifx [1]{%
 \ifx #1\expandafter \@firstoftwo
 \else \expandafter \@secondoftwo
 \fi
}%
\providecommand \natexlab [1]{#1}%
\providecommand \enquote  [1]{``#1''}%
\providecommand \bibnamefont  [1]{#1}%
\providecommand \bibfnamefont [1]{#1}%
\providecommand \citenamefont [1]{#1}%
\providecommand \href@noop [0]{\@secondoftwo}%
\providecommand \href [0]{\begingroup \@sanitize@url \@href}%
\providecommand \@href[1]{\@@startlink{#1}\@@href}%
\providecommand \@@href[1]{\endgroup#1\@@endlink}%
\providecommand \@sanitize@url [0]{\catcode `\\12\catcode `\$12\catcode
  `\&12\catcode `\#12\catcode `\^12\catcode `\_12\catcode `\%12\relax}%
\providecommand \@@startlink[1]{}%
\providecommand \@@endlink[0]{}%
\providecommand \url  [0]{\begingroup\@sanitize@url \@url }%
\providecommand \@url [1]{\endgroup\@href {#1}{\urlprefix }}%
\providecommand \urlprefix  [0]{URL }%
\providecommand \Eprint [0]{\href }%
\providecommand \doibase [0]{http://dx.doi.org/}%
\providecommand \selectlanguage [0]{\@gobble}%
\providecommand \bibinfo  [0]{\@secondoftwo}%
\providecommand \bibfield  [0]{\@secondoftwo}%
\providecommand \translation [1]{[#1]}%
\providecommand \BibitemOpen [0]{}%
\providecommand \bibitemStop [0]{}%
\providecommand \bibitemNoStop [0]{.\EOS\space}%
\providecommand \EOS [0]{\spacefactor3000\relax}%
\providecommand \BibitemShut  [1]{\csname bibitem#1\endcsname}%
\let\auto@bib@innerbib\@empty
%</preamble>
\bibitem [{\citenamefont {Hopf}(1952)}]{hopf_statistical_1952}%
  \BibitemOpen
  \bibfield  {author} {\bibinfo {author} {\bibfnamefont {E.}~\bibnamefont
  {Hopf}},\ }\href {https://www.jstor.org/stable/24900259} {\bibfield
  {journal} {\bibinfo  {journal} {Journal of Rational Mechanics and Analysis}\
  }\textbf {\bibinfo {volume} {1}},\ \bibinfo {pages} {87} (\bibinfo {year}
  {1952})}\BibitemShut {NoStop}%
\bibitem [{\citenamefont {Lee}\ and\ \citenamefont
  {Yang}(1952)}]{lee_statistical_1952}%
  \BibitemOpen
  \bibfield  {author} {\bibinfo {author} {\bibfnamefont {T.~D.}\ \bibnamefont
  {Lee}}\ and\ \bibinfo {author} {\bibfnamefont {C.~N.}\ \bibnamefont {Yang}},\
  }\href {\doibase 10.1103/PhysRev.87.410} {\bibfield  {journal} {\bibinfo
  {journal} {Phys. Rev.}\ }\textbf {\bibinfo {volume} {87}},\ \bibinfo {pages}
  {410} (\bibinfo {year} {1952})}\BibitemShut {NoStop}%
\bibitem [{\citenamefont {Kraichnan}(1967)}]{kraichnan_inertial_1967}%
  \BibitemOpen
  \bibfield  {author} {\bibinfo {author} {\bibfnamefont {R.~H.}\ \bibnamefont
  {Kraichnan}},\ }\href {\doibase 10.1063/1.1762301} {\bibfield  {journal}
  {\bibinfo  {journal} {The Physics of Fluids}\ }\textbf {\bibinfo {volume}
  {10}},\ \bibinfo {pages} {1417} (\bibinfo {year} {1967})}\BibitemShut
  {NoStop}%
\bibitem [{\citenamefont {Kraichnan}(1973)}]{kraichnan_helical_1973}%
  \BibitemOpen
  \bibfield  {author} {\bibinfo {author} {\bibfnamefont {R.~H.}\ \bibnamefont
  {Kraichnan}},\ }\href {\doibase 10.1017/S0022112073001837} {\bibfield
  {journal} {\bibinfo  {journal} {Journal of Fluid Mechanics}\ }\textbf
  {\bibinfo {volume} {59}},\ \bibinfo {pages} {745} (\bibinfo {year}
  {1973})}\BibitemShut {NoStop}%
\bibitem [{\citenamefont {Cichowlas}\ and\ \citenamefont
  {Brachet}(2005)}]{cichowlas_evolution_2005}%
  \BibitemOpen
  \bibfield  {author} {\bibinfo {author} {\bibfnamefont {C.}~\bibnamefont
  {Cichowlas}}\ and\ \bibinfo {author} {\bibfnamefont {M.-E.}\ \bibnamefont
  {Brachet}},\ }\href {\doibase https://doi.org/10.1016/j.fluiddyn.2004.09.005}
  {\bibfield  {journal} {\bibinfo  {journal} {Fluid Dynamics Research}\
  }\textbf {\bibinfo {volume} {36}},\ \bibinfo {pages} {239} (\bibinfo {year}
  {2005})}\BibitemShut {NoStop}%
\bibitem [{\citenamefont {Cichowlas}\ \emph {et~al.}(2005)\citenamefont
  {Cichowlas}, \citenamefont {Bonaïti}, \citenamefont {Debbasch},\ and\
  \citenamefont {Brachet}}]{cichowlas_effective_2005}%
  \BibitemOpen
  \bibfield  {author} {\bibinfo {author} {\bibfnamefont {C.}~\bibnamefont
  {Cichowlas}}, \bibinfo {author} {\bibfnamefont {P.}~\bibnamefont {Bonaïti}},
  \bibinfo {author} {\bibfnamefont {F.}~\bibnamefont {Debbasch}}, \ and\
  \bibinfo {author} {\bibfnamefont {M.}~\bibnamefont {Brachet}},\ }\href
  {\doibase 10.1103/PhysRevLett.95.264502} {\bibfield  {journal} {\bibinfo
  {journal} {Phys. Rev. Lett.}\ }\textbf {\bibinfo {volume} {95}},\ \bibinfo
  {pages} {264502} (\bibinfo {year} {2005})}\BibitemShut {NoStop}%
\bibitem [{\citenamefont {Krstulovic}\ and\ \citenamefont
  {Brachet}(2008)}]{krstulovic_two-fluid_2008}%
  \BibitemOpen
  \bibfield  {author} {\bibinfo {author} {\bibfnamefont {G.}~\bibnamefont
  {Krstulovic}}\ and\ \bibinfo {author} {\bibfnamefont {M.-E.}\ \bibnamefont
  {Brachet}},\ }\href {\doibase 10.1016/j.physd.2007.11.008} {\bibfield
  {journal} {\bibinfo  {journal} {Physica D: Nonlinear Phenomena}\ }\textbf
  {\bibinfo {volume} {237}},\ \bibinfo {pages} {2015} (\bibinfo {year}
  {2008})}\BibitemShut {NoStop}%
\bibitem [{\citenamefont {Hopf}(1950)}]{hopf_partial_1950}%
  \BibitemOpen
  \bibfield  {author} {\bibinfo {author} {\bibfnamefont {E.}~\bibnamefont
  {Hopf}},\ }\href {\doibase 10.1002/CPA.3160030302} {\bibfield  {journal}
  {\bibinfo  {journal} {Comm. Pure Appl. Math.}\ }\textbf {\bibinfo {volume}
  {3}},\ \bibinfo {pages} {201} (\bibinfo {year} {1950})}\BibitemShut {NoStop}%
\bibitem [{\citenamefont {Murugan}\ \emph {et~al.}(2021)\citenamefont
  {Murugan}, \citenamefont {Kumar}, \citenamefont {Bhattacharjee},\ and\
  \citenamefont {Ray}}]{murugan_many-body_2021}%
  \BibitemOpen
  \bibfield  {author} {\bibinfo {author} {\bibfnamefont {S.~D.}\ \bibnamefont
  {Murugan}}, \bibinfo {author} {\bibfnamefont {D.}~\bibnamefont {Kumar}},
  \bibinfo {author} {\bibfnamefont {S.}~\bibnamefont {Bhattacharjee}}, \ and\
  \bibinfo {author} {\bibfnamefont {S.~S.}\ \bibnamefont {Ray}},\ }\href
  {\doibase 10.1103/PhysRevLett.127.124501} {\bibfield  {journal} {\bibinfo
  {journal} {Phys. Rev. Lett.}\ }\textbf {\bibinfo {volume} {127}},\ \bibinfo
  {pages} {124501} (\bibinfo {year} {2021})}\BibitemShut {NoStop}%
\bibitem [{\citenamefont {Rampf}\ \emph {et~al.}(2022)\citenamefont {Rampf},
  \citenamefont {Frisch},\ and\ \citenamefont {Hahn}}]{rampf_eye_2022}%
  \BibitemOpen
  \bibfield  {author} {\bibinfo {author} {\bibfnamefont {C.}~\bibnamefont
  {Rampf}}, \bibinfo {author} {\bibfnamefont {U.}~\bibnamefont {Frisch}}, \
  and\ \bibinfo {author} {\bibfnamefont {O.}~\bibnamefont {Hahn}},\ }\href
  {\doibase 10.48550/ARXIV.2207.12416} {\bibfield  {journal} {\bibinfo
  {journal} {arXiv e-prints}\ ,\ \bibinfo {pages} {arXiv:2207.12416}} (\bibinfo
  {year} {2022})}\BibitemShut {NoStop}%
\bibitem [{\citenamefont {Bec}\ and\ \citenamefont
  {Khanin}(2007)}]{bec_burgers_2007}%
  \BibitemOpen
  \bibfield  {author} {\bibinfo {author} {\bibfnamefont {J.}~\bibnamefont
  {Bec}}\ and\ \bibinfo {author} {\bibfnamefont {K.}~\bibnamefont {Khanin}},\
  }\href {\doibase https://doi.org/10.1016/j.physrep.2007.04.002} {\bibfield
  {journal} {\bibinfo  {journal} {Phys. Rep.}\ }\textbf {\bibinfo {volume}
  {447}},\ \bibinfo {pages} {1 } (\bibinfo {year} {2007})}\BibitemShut
  {NoStop}%
\bibitem [{\citenamefont {{Rose, H. A.}}\ and\ \citenamefont {{Sulem, P.
  L.}}(1978)}]{rose_ha_fully_1978}%
  \BibitemOpen
  \bibfield  {author} {\bibinfo {author} {\bibnamefont {{Rose, H. A.}}}\ and\
  \bibinfo {author} {\bibnamefont {{Sulem, P. L.}}},\ }\href {\doibase
  10.1051/jphys:01978003905044100} {\bibfield  {journal} {\bibinfo  {journal}
  {J. Phys. France}\ }\textbf {\bibinfo {volume} {39}},\ \bibinfo {pages} {441}
  (\bibinfo {year} {1978})}\BibitemShut {NoStop}%
\bibitem [{\citenamefont {Orszag}(1970)}]{orszag_analytical_1970}%
  \BibitemOpen
  \bibfield  {author} {\bibinfo {author} {\bibfnamefont {S.~A.}\ \bibnamefont
  {Orszag}},\ }\href {\doibase 10.1017/S0022112070000642} {\bibfield  {journal}
  {\bibinfo  {journal} {Journal of Fluid Mechanics}\ }\textbf {\bibinfo
  {volume} {41}},\ \bibinfo {pages} {363} (\bibinfo {year} {1970})}\BibitemShut
  {NoStop}%
\bibitem [{\citenamefont {Frisch}(1995)}]{frisch_turbulence_1995}%
  \BibitemOpen
  \bibfield  {author} {\bibinfo {author} {\bibfnamefont {U.}~\bibnamefont
  {Frisch}},\ }\href {\doibase 10.1017/CBO9781139170666} {{\selectlanguage
  {english}\emph {\bibinfo {title} {Turbulence: {The} {Legacy} of {A}. {N}.
  {Kolmogorov}}}}}\ (\bibinfo  {publisher} {Cambridge University Press,
  Cambridge, The United Kingdom},\ \bibinfo {year} {1995})\BibitemShut
  {NoStop}%
\bibitem [{\citenamefont {Kraichnan}(1959)}]{kraichnan_structure_1959}%
  \BibitemOpen
  \bibfield  {author} {\bibinfo {author} {\bibfnamefont {R.~H.}\ \bibnamefont
  {Kraichnan}},\ }\href {\doibase 10.1017/S0022112059000362} {\bibfield
  {journal} {\bibinfo  {journal} {Journal of Fluid Mechanics}\ }\textbf
  {\bibinfo {volume} {5}},\ \bibinfo {pages} {497} (\bibinfo {year}
  {1959})}\BibitemShut {NoStop}%
\bibitem [{\citenamefont {Kolmogorov}\ \emph {et~al.}(1991)\citenamefont
  {Kolmogorov}, \citenamefont {Levin}, \citenamefont {Hunt}, \citenamefont
  {Phillips},\ and\ \citenamefont {Williams}}]{kolmogorov_local_1991}%
  \BibitemOpen
  \bibfield  {author} {\bibinfo {author} {\bibfnamefont {A.~N.}\ \bibnamefont
  {Kolmogorov}}, \bibinfo {author} {\bibfnamefont {V.}~\bibnamefont {Levin}},
  \bibinfo {author} {\bibfnamefont {J.~C.~R.}\ \bibnamefont {Hunt}}, \bibinfo
  {author} {\bibfnamefont {O.~M.}\ \bibnamefont {Phillips}}, \ and\ \bibinfo
  {author} {\bibfnamefont {D.}~\bibnamefont {Williams}},\ }\href {\doibase
  10.1098/rspa.1991.0075} {\bibfield  {journal} {\bibinfo  {journal}
  {Proceedings of the Royal Society of London. Series A: Mathematical and
  Physical Sciences}\ }\textbf {\bibinfo {volume} {434}},\ \bibinfo {pages} {9}
  (\bibinfo {year} {1991})}\BibitemShut {NoStop}%
\bibitem [{\citenamefont {L'vov}\ \emph {et~al.}(2002)\citenamefont {L'vov},
  \citenamefont {Pomyalov},\ and\ \citenamefont
  {Procaccia}}]{l'vov_quasi-gaussian_2002}%
  \BibitemOpen
  \bibfield  {author} {\bibinfo {author} {\bibfnamefont {V.~S.}\ \bibnamefont
  {L'vov}}, \bibinfo {author} {\bibfnamefont {A.}~\bibnamefont {Pomyalov}}, \
  and\ \bibinfo {author} {\bibfnamefont {I.}~\bibnamefont {Procaccia}},\ }\href
  {\doibase 10.1103/PhysRevLett.89.064501} {\bibfield  {journal} {\bibinfo
  {journal} {Phys. Rev. Lett.}\ }\textbf {\bibinfo {volume} {89}},\ \bibinfo
  {pages} {064501} (\bibinfo {year} {2002})}\BibitemShut {NoStop}%
\bibitem [{\citenamefont {Frisch}\ \emph {et~al.}(2012)\citenamefont {Frisch},
  \citenamefont {Pomyalov}, \citenamefont {Procaccia},\ and\ \citenamefont
  {Ray}}]{frisch_turbulence_2012}%
  \BibitemOpen
  \bibfield  {author} {\bibinfo {author} {\bibfnamefont {U.}~\bibnamefont
  {Frisch}}, \bibinfo {author} {\bibfnamefont {A.}~\bibnamefont {Pomyalov}},
  \bibinfo {author} {\bibfnamefont {I.}~\bibnamefont {Procaccia}}, \ and\
  \bibinfo {author} {\bibfnamefont {S.~S.}\ \bibnamefont {Ray}},\ }\href
  {\doibase 10.1103/PhysRevLett.108.074501} {\bibfield  {journal} {\bibinfo
  {journal} {Phys. Rev. Lett.}\ }\textbf {\bibinfo {volume} {108}},\ \bibinfo
  {pages} {074501} (\bibinfo {year} {2012})}\BibitemShut {NoStop}%
\bibitem [{\citenamefont {Lanotte}\ \emph {et~al.}(2015)\citenamefont
  {Lanotte}, \citenamefont {Benzi}, \citenamefont {Malapaka}, \citenamefont
  {Toschi},\ and\ \citenamefont {Biferale}}]{lanotte_turbulence_2015}%
  \BibitemOpen
  \bibfield  {author} {\bibinfo {author} {\bibfnamefont {A.~S.}\ \bibnamefont
  {Lanotte}}, \bibinfo {author} {\bibfnamefont {R.}~\bibnamefont {Benzi}},
  \bibinfo {author} {\bibfnamefont {S.~K.}\ \bibnamefont {Malapaka}}, \bibinfo
  {author} {\bibfnamefont {F.}~\bibnamefont {Toschi}}, \ and\ \bibinfo {author}
  {\bibfnamefont {L.}~\bibnamefont {Biferale}},\ }\href {\doibase
  10.1103/PhysRevLett.115.264502} {\bibfield  {journal} {\bibinfo  {journal}
  {Phys. Rev. Lett.}\ }\textbf {\bibinfo {volume} {115}},\ \bibinfo {pages}
  {264502} (\bibinfo {year} {2015})}\BibitemShut {NoStop}%
\bibitem [{\citenamefont {Lanotte}\ \emph {et~al.}(2016)\citenamefont
  {Lanotte}, \citenamefont {Malapaka},\ and\ \citenamefont
  {Biferale}}]{lanotte_vortex_2016}%
  \BibitemOpen
  \bibfield  {author} {\bibinfo {author} {\bibfnamefont {A.~S.}\ \bibnamefont
  {Lanotte}}, \bibinfo {author} {\bibfnamefont {S.~K.}\ \bibnamefont
  {Malapaka}}, \ and\ \bibinfo {author} {\bibfnamefont {L.}~\bibnamefont
  {Biferale}},\ }\href {\doibase 10.1140/epje/i2016-16049-x} {\bibfield
  {journal} {\bibinfo  {journal} {Eur. Phys. J. E}\ }\textbf {\bibinfo {volume}
  {39}},\ \bibinfo {pages} {49} (\bibinfo {year} {2016})}\BibitemShut {NoStop}%
\bibitem [{\citenamefont {Buzzicotti}\ \emph
  {et~al.}(2016{\natexlab{a}})\citenamefont {Buzzicotti}, \citenamefont
  {Biferale}, \citenamefont {Frisch},\ and\ \citenamefont
  {Ray}}]{buzzicotti_intermittency_2016}%
  \BibitemOpen
  \bibfield  {author} {\bibinfo {author} {\bibfnamefont {M.}~\bibnamefont
  {Buzzicotti}}, \bibinfo {author} {\bibfnamefont {L.}~\bibnamefont
  {Biferale}}, \bibinfo {author} {\bibfnamefont {U.}~\bibnamefont {Frisch}}, \
  and\ \bibinfo {author} {\bibfnamefont {S.~S.}\ \bibnamefont {Ray}},\ }\href
  {\doibase 10.1103/PhysRevE.93.033109} {\bibfield  {journal} {\bibinfo
  {journal} {Phys. Rev. E}\ }\textbf {\bibinfo {volume} {93}},\ \bibinfo
  {pages} {033109} (\bibinfo {year} {2016}{\natexlab{a}})}\BibitemShut
  {NoStop}%
\bibitem [{\citenamefont {Buzzicotti}\ \emph
  {et~al.}(2016{\natexlab{b}})\citenamefont {Buzzicotti}, \citenamefont
  {Bhatnagar}, \citenamefont {Biferale}, \citenamefont {Lanotte},\ and\
  \citenamefont {Ray}}]{buzzicotti_lagrangian_2016}%
  \BibitemOpen
  \bibfield  {author} {\bibinfo {author} {\bibfnamefont {M.}~\bibnamefont
  {Buzzicotti}}, \bibinfo {author} {\bibfnamefont {A.}~\bibnamefont
  {Bhatnagar}}, \bibinfo {author} {\bibfnamefont {L.}~\bibnamefont {Biferale}},
  \bibinfo {author} {\bibfnamefont {A.~S.}\ \bibnamefont {Lanotte}}, \ and\
  \bibinfo {author} {\bibfnamefont {S.~S.}\ \bibnamefont {Ray}},\ }\href
  {\doibase 10.1088/1367-2630/18/11/113047} {\bibfield  {journal} {\bibinfo
  {journal} {New J. Phys.}\ }\textbf {\bibinfo {volume} {18}},\ \bibinfo
  {pages} {113047} (\bibinfo {year} {2016}{\natexlab{b}})}\BibitemShut
  {NoStop}%
\bibitem [{\citenamefont {Ray}(2018)}]{ray_non-intermittent_2018}%
  \BibitemOpen
  \bibfield  {author} {\bibinfo {author} {\bibfnamefont {S.~S.}\ \bibnamefont
  {Ray}},\ }\href {\doibase 10.1103/PhysRevFluids.3.072601} {\bibfield
  {journal} {\bibinfo  {journal} {Phys. Rev. Fluids}\ }\textbf {\bibinfo
  {volume} {3}},\ \bibinfo {pages} {072601} (\bibinfo {year}
  {2018})}\BibitemShut {NoStop}%
\bibitem [{\citenamefont {Tom}\ and\ \citenamefont
  {Ray}(2017)}]{tom_revisiting_2017}%
  \BibitemOpen
  \bibfield  {author} {\bibinfo {author} {\bibfnamefont {R.}~\bibnamefont
  {Tom}}\ and\ \bibinfo {author} {\bibfnamefont {S.~S.}\ \bibnamefont {Ray}},\
  }\href {\doibase 10.1209/0295-5075/120/34002} {\bibfield  {journal} {\bibinfo
   {journal} {Europhys. Lett.}\ }\textbf {\bibinfo {volume} {120}},\ \bibinfo
  {pages} {34002} (\bibinfo {year} {2017})}\BibitemShut {NoStop}%
\bibitem [{\citenamefont {Picardo}\ \emph {et~al.}(2020)\citenamefont
  {Picardo}, \citenamefont {Bhatnagar},\ and\ \citenamefont
  {Ray}}]{picardo_lagrangian_2020}%
  \BibitemOpen
  \bibfield  {author} {\bibinfo {author} {\bibfnamefont {J.~R.}\ \bibnamefont
  {Picardo}}, \bibinfo {author} {\bibfnamefont {A.}~\bibnamefont {Bhatnagar}},
  \ and\ \bibinfo {author} {\bibfnamefont {S.~S.}\ \bibnamefont {Ray}},\ }\href
  {\doibase 10.1103/PhysRevFluids.5.042601} {\bibfield  {journal} {\bibinfo
  {journal} {Phys. Rev. Fluids}\ }\textbf {\bibinfo {volume} {5}},\ \bibinfo
  {pages} {042601} (\bibinfo {year} {2020})}\BibitemShut {NoStop}%
\bibitem [{\citenamefont {Frisch}\ \emph {et~al.}(2008)\citenamefont {Frisch},
  \citenamefont {Kurien}, \citenamefont {Pandit}, \citenamefont {Pauls},
  \citenamefont {Ray}, \citenamefont {Wirth},\ and\ \citenamefont
  {Zhu}}]{frisch_hyperviscosity_2008}%
  \BibitemOpen
  \bibfield  {author} {\bibinfo {author} {\bibfnamefont {U.}~\bibnamefont
  {Frisch}}, \bibinfo {author} {\bibfnamefont {S.}~\bibnamefont {Kurien}},
  \bibinfo {author} {\bibfnamefont {R.}~\bibnamefont {Pandit}}, \bibinfo
  {author} {\bibfnamefont {W.}~\bibnamefont {Pauls}}, \bibinfo {author}
  {\bibfnamefont {S.~S.}\ \bibnamefont {Ray}}, \bibinfo {author} {\bibfnamefont
  {A.}~\bibnamefont {Wirth}}, \ and\ \bibinfo {author} {\bibfnamefont {J.-Z.}\
  \bibnamefont {Zhu}},\ }\href {\doibase 10.1103/PhysRevLett.101.144501}
  {\bibfield  {journal} {\bibinfo  {journal} {Phys. Rev. Lett.}\ }\textbf
  {\bibinfo {volume} {101}},\ \bibinfo {pages} {144501} (\bibinfo {year}
  {2008})}\BibitemShut {NoStop}%
\bibitem [{\citenamefont {Frisch}\ \emph {et~al.}(2013)\citenamefont {Frisch},
  \citenamefont {Ray}, \citenamefont {Sahoo}, \citenamefont {Banerjee},\ and\
  \citenamefont {Pandit}}]{frisch_real-space_2013}%
  \BibitemOpen
  \bibfield  {author} {\bibinfo {author} {\bibfnamefont {U.}~\bibnamefont
  {Frisch}}, \bibinfo {author} {\bibfnamefont {S.~S.}\ \bibnamefont {Ray}},
  \bibinfo {author} {\bibfnamefont {G.}~\bibnamefont {Sahoo}}, \bibinfo
  {author} {\bibfnamefont {D.}~\bibnamefont {Banerjee}}, \ and\ \bibinfo
  {author} {\bibfnamefont {R.}~\bibnamefont {Pandit}},\ }\href {\doibase
  10.1103/PhysRevLett.110.064501} {\bibfield  {journal} {\bibinfo  {journal}
  {Phys. Rev. Lett.}\ }\textbf {\bibinfo {volume} {110}},\ \bibinfo {pages}
  {064501} (\bibinfo {year} {2013})}\BibitemShut {NoStop}%
\bibitem [{\citenamefont {Banerjee}\ and\ \citenamefont
  {Ray}(2014)}]{banerjee_transition_2014}%
  \BibitemOpen
  \bibfield  {author} {\bibinfo {author} {\bibfnamefont {D.}~\bibnamefont
  {Banerjee}}\ and\ \bibinfo {author} {\bibfnamefont {S.~S.}\ \bibnamefont
  {Ray}},\ }\href {\doibase 10.1103/PhysRevE.90.041001} {\bibfield  {journal}
  {\bibinfo  {journal} {Phys. Rev. E}\ }\textbf {\bibinfo {volume} {90}},\
  \bibinfo {pages} {041001} (\bibinfo {year} {2014})}\BibitemShut {NoStop}%
\bibitem [{\citenamefont {Bandak}\ \emph {et~al.}(2022)\citenamefont {Bandak},
  \citenamefont {Goldenfeld}, \citenamefont {Mailybaev},\ and\ \citenamefont
  {Eyink}}]{bandak_dissipation-range_2022}%
  \BibitemOpen
  \bibfield  {author} {\bibinfo {author} {\bibfnamefont {D.}~\bibnamefont
  {Bandak}}, \bibinfo {author} {\bibfnamefont {N.}~\bibnamefont {Goldenfeld}},
  \bibinfo {author} {\bibfnamefont {A.~A.}\ \bibnamefont {Mailybaev}}, \ and\
  \bibinfo {author} {\bibfnamefont {G.}~\bibnamefont {Eyink}},\ }\href
  {\doibase 10.1103/PhysRevE.105.065113} {\bibfield  {journal} {\bibinfo
  {journal} {Phys. Rev. E}\ }\textbf {\bibinfo {volume} {105}},\ \bibinfo
  {pages} {065113} (\bibinfo {year} {2022})}\BibitemShut {NoStop}%
\bibitem [{\citenamefont {Fournier}\ and\ \citenamefont
  {Frisch}(1978)}]{fournier_d-dimensional_1978}%
  \BibitemOpen
  \bibfield  {author} {\bibinfo {author} {\bibfnamefont {J.-D.}\ \bibnamefont
  {Fournier}}\ and\ \bibinfo {author} {\bibfnamefont {U.}~\bibnamefont
  {Frisch}},\ }\href {\doibase 10.1103/PhysRevA.17.747} {\bibfield  {journal}
  {\bibinfo  {journal} {Phys. Rev. A}\ }\textbf {\bibinfo {volume} {17}},\
  \bibinfo {pages} {747} (\bibinfo {year} {1978})}\BibitemShut {NoStop}%
\bibitem [{\citenamefont {Celani}\ \emph {et~al.}(2010)\citenamefont {Celani},
  \citenamefont {Musacchio},\ and\ \citenamefont
  {Vincenzi}}]{celani_turbulence_2010}%
  \BibitemOpen
  \bibfield  {author} {\bibinfo {author} {\bibfnamefont {A.}~\bibnamefont
  {Celani}}, \bibinfo {author} {\bibfnamefont {S.}~\bibnamefont {Musacchio}}, \
  and\ \bibinfo {author} {\bibfnamefont {D.}~\bibnamefont {Vincenzi}},\ }\href
  {\doibase 10.1103/PhysRevLett.104.184506} {\bibfield  {journal} {\bibinfo
  {journal} {Phys. Rev. Lett.}\ }\textbf {\bibinfo {volume} {104}},\ \bibinfo
  {pages} {184506} (\bibinfo {year} {2010})}\BibitemShut {NoStop}%
\bibitem [{\citenamefont {{Leray}}\ and\ \citenamefont
  {{Terrell}}(2016)}]{leray_motion_2016}%
  \BibitemOpen
  \bibfield  {author} {\bibinfo {author} {\bibfnamefont {J.}~\bibnamefont
  {{Leray}}}\ and\ \bibinfo {author} {\bibfnamefont {R.}~\bibnamefont
  {{Terrell}}},\ }\href@noop {} {\bibfield  {journal} {\bibinfo  {journal}
  {arXiv e-prints}\ ,\ \bibinfo {eid} {arXiv:1604.02484}} (\bibinfo {year}
  {2016})},\ \Eprint {http://arxiv.org/abs/1604.02484} {arXiv:1604.02484
  [math.HO]} \BibitemShut {NoStop}%
\bibitem [{\citenamefont {Onsager}(1949)}]{onsager_statistical_1949}%
  \BibitemOpen
  \bibfield  {author} {\bibinfo {author} {\bibfnamefont {L.}~\bibnamefont
  {Onsager}},\ }\href {\doibase 10.1007/BF02780991} {\bibfield  {journal}
  {\bibinfo  {journal} {Nuovo Cim}\ }\textbf {\bibinfo {volume} {6}},\ \bibinfo
  {pages} {279} (\bibinfo {year} {1949})}\BibitemShut {NoStop}%
\bibitem [{\citenamefont {Gibbon}(2008)}]{gibbon_three-dimensional_2008}%
  \BibitemOpen
  \bibfield  {author} {\bibinfo {author} {\bibfnamefont {J.~D.}\ \bibnamefont
  {Gibbon}},\ }\href {\doibase 10.1016/j.physd.2007.10.014} {\bibfield
  {journal} {\bibinfo  {journal} {Physica D: Nonlinear Phenomena}\ }\bibinfo
  {series} {Euler {Equations}: 250 {Years} {On}},\ \textbf {\bibinfo {volume}
  {237}},\ \bibinfo {pages} {1894} (\bibinfo {year} {2008})}\BibitemShut
  {NoStop}%
\bibitem [{\citenamefont {Gibbon}\ \emph {et~al.}(2008)\citenamefont {Gibbon},
  \citenamefont {Bustamante},\ and\ \citenamefont
  {Kerr}}]{gibbon_three-dimensional_2008-1}%
  \BibitemOpen
  \bibfield  {author} {\bibinfo {author} {\bibfnamefont {J.~D.}\ \bibnamefont
  {Gibbon}}, \bibinfo {author} {\bibfnamefont {M.}~\bibnamefont {Bustamante}},
  \ and\ \bibinfo {author} {\bibfnamefont {R.~M.}\ \bibnamefont {Kerr}},\
  }\href {\doibase 10.1088/0951-7715/21/8/T02} {\bibfield  {journal} {\bibinfo
  {journal} {Nonlinearity}\ }\textbf {\bibinfo {volume} {21}},\ \bibinfo
  {pages} {T123} (\bibinfo {year} {2008})}\BibitemShut {NoStop}%
\bibitem [{\citenamefont {Frisch}\ \emph {et~al.}(2002)\citenamefont {Frisch},
  \citenamefont {Matsumoto},\ and\ \citenamefont
  {Bec}}]{frisch_singularities_2002}%
  \BibitemOpen
  \bibfield  {author} {\bibinfo {author} {\bibfnamefont {U.}~\bibnamefont
  {Frisch}}, \bibinfo {author} {\bibfnamefont {T.}~\bibnamefont {Matsumoto}}, \
  and\ \bibinfo {author} {\bibfnamefont {J.}~\bibnamefont {Bec}},\ }\href
  {\doibase 10.1023/A:1027308602344} {\bibfield  {journal} {\bibinfo  {journal}
  {Journal of Statistical Physics}\ }\textbf {\bibinfo {volume} {113}}
  (\bibinfo {year} {2002}),\ 10.1023/A:1027308602344}\BibitemShut {NoStop}%
\bibitem [{\citenamefont {Brachet}\ \emph {et~al.}(1983)\citenamefont
  {Brachet}, \citenamefont {Meiron}, \citenamefont {Orszag}, \citenamefont
  {Nickel}, \citenamefont {Morf},\ and\ \citenamefont
  {Frisch}}]{brachet_small-scale_1983}%
  \BibitemOpen
  \bibfield  {author} {\bibinfo {author} {\bibfnamefont {M.~E.}\ \bibnamefont
  {Brachet}}, \bibinfo {author} {\bibfnamefont {D.~I.}\ \bibnamefont {Meiron}},
  \bibinfo {author} {\bibfnamefont {S.~A.}\ \bibnamefont {Orszag}}, \bibinfo
  {author} {\bibfnamefont {B.~G.}\ \bibnamefont {Nickel}}, \bibinfo {author}
  {\bibfnamefont {R.~H.}\ \bibnamefont {Morf}}, \ and\ \bibinfo {author}
  {\bibfnamefont {U.}~\bibnamefont {Frisch}},\ }\href {\doibase
  10.1017/S0022112083001159} {\bibfield  {journal} {\bibinfo  {journal}
  {Journal of Fluid Mechanics}\ }\textbf {\bibinfo {volume} {130}},\ \bibinfo
  {pages} {411} (\bibinfo {year} {1983})}\BibitemShut {NoStop}%
\bibitem [{\citenamefont {Brachet}\ \emph {et~al.}(1992)\citenamefont
  {Brachet}, \citenamefont {Meneguzzi}, \citenamefont {Vincent}, \citenamefont
  {Politano},\ and\ \citenamefont {Sulem}}]{brachet_numerical_1992}%
  \BibitemOpen
  \bibfield  {author} {\bibinfo {author} {\bibfnamefont {M.~E.}\ \bibnamefont
  {Brachet}}, \bibinfo {author} {\bibfnamefont {M.}~\bibnamefont {Meneguzzi}},
  \bibinfo {author} {\bibfnamefont {A.}~\bibnamefont {Vincent}}, \bibinfo
  {author} {\bibfnamefont {H.}~\bibnamefont {Politano}}, \ and\ \bibinfo
  {author} {\bibfnamefont {P.~L.}\ \bibnamefont {Sulem}},\ }\href {\doibase
  10.1063/1.858513} {\bibfield  {journal} {\bibinfo  {journal} {Physics of
  Fluids A: Fluid Dynamics}\ }\textbf {\bibinfo {volume} {4}},\ \bibinfo
  {pages} {2845} (\bibinfo {year} {1992})}\BibitemShut {NoStop}%
\bibitem [{\citenamefont {Boratav}\ \emph {et~al.}(1992)\citenamefont
  {Boratav}, \citenamefont {Pelz},\ and\ \citenamefont
  {Zabusky}}]{boratav_reconnection_1992}%
  \BibitemOpen
  \bibfield  {author} {\bibinfo {author} {\bibfnamefont {O.~N.}\ \bibnamefont
  {Boratav}}, \bibinfo {author} {\bibfnamefont {R.~B.}\ \bibnamefont {Pelz}}, \
  and\ \bibinfo {author} {\bibfnamefont {N.~J.}\ \bibnamefont {Zabusky}},\
  }\href {\doibase 10.1063/1.858329} {\bibfield  {journal} {\bibinfo  {journal}
  {Physics of Fluids A: Fluid Dynamics}\ }\textbf {\bibinfo {volume} {4}},\
  \bibinfo {pages} {581} (\bibinfo {year} {1992})}\BibitemShut {NoStop}%
\bibitem [{\citenamefont {Kerr}(1993)}]{kerr_evidence_1993}%
  \BibitemOpen
  \bibfield  {author} {\bibinfo {author} {\bibfnamefont {R.~M.}\ \bibnamefont
  {Kerr}},\ }\href {\doibase 10.1063/1.858849} {\bibfield  {journal} {\bibinfo
  {journal} {Physics of Fluids A: Fluid Dynamics}\ }\textbf {\bibinfo {volume}
  {5}},\ \bibinfo {pages} {1725} (\bibinfo {year} {1993})}\BibitemShut
  {NoStop}%
\bibitem [{\citenamefont {Shelley}\ \emph {et~al.}(1993)\citenamefont
  {Shelley}, \citenamefont {Meiron},\ and\ \citenamefont
  {Orszag}}]{shelley_dynamical_1993}%
  \BibitemOpen
  \bibfield  {author} {\bibinfo {author} {\bibfnamefont {M.~J.}\ \bibnamefont
  {Shelley}}, \bibinfo {author} {\bibfnamefont {D.~I.}\ \bibnamefont {Meiron}},
  \ and\ \bibinfo {author} {\bibfnamefont {S.~A.}\ \bibnamefont {Orszag}},\
  }\href {\doibase 10.1017/S0022112093000291} {\bibfield  {journal} {\bibinfo
  {journal} {J. Fluid Mech.}\ }\textbf {\bibinfo {volume} {246}},\ \bibinfo
  {pages} {613} (\bibinfo {year} {1993})}\BibitemShut {NoStop}%
\bibitem [{\citenamefont {Boratav}\ and\ \citenamefont
  {Pelz}(1994)}]{boratav_direct_1994}%
  \BibitemOpen
  \bibfield  {author} {\bibinfo {author} {\bibfnamefont {O.~N.}\ \bibnamefont
  {Boratav}}\ and\ \bibinfo {author} {\bibfnamefont {R.~B.}\ \bibnamefont
  {Pelz}},\ }\href {\doibase 10.1063/1.868166} {\bibfield  {journal} {\bibinfo
  {journal} {Physics of Fluids}\ }\textbf {\bibinfo {volume} {6}},\ \bibinfo
  {pages} {2757} (\bibinfo {year} {1994})}\BibitemShut {NoStop}%
\bibitem [{\citenamefont {Kerr}(2005)}]{kerr_velocity_2005}%
  \BibitemOpen
  \bibfield  {author} {\bibinfo {author} {\bibfnamefont {R.~M.}\ \bibnamefont
  {Kerr}},\ }\href {\doibase 10.1063/1.1905183} {\bibfield  {journal} {\bibinfo
   {journal} {Physics of Fluids}\ }\textbf {\bibinfo {volume} {17}},\ \bibinfo
  {pages} {075103} (\bibinfo {year} {2005})}\BibitemShut {NoStop}%
\bibitem [{\citenamefont {Pelz}\ and\ \citenamefont
  {Ohkitani}(2005)}]{pelz_linearly_2005}%
  \BibitemOpen
  \bibfield  {author} {\bibinfo {author} {\bibfnamefont {R.}~\bibnamefont
  {Pelz}}\ and\ \bibinfo {author} {\bibfnamefont {K.}~\bibnamefont
  {Ohkitani}},\ }\href {\doibase 10.1016/j.fluiddyn.2004.10.005} {\bibfield
  {journal} {\bibinfo  {journal} {Fluid Dynamics Research}\ }\textbf {\bibinfo
  {volume} {36}},\ \bibinfo {pages} {193} (\bibinfo {year} {2005})}\BibitemShut
  {NoStop}%
\bibitem [{\citenamefont {Hou}\ and\ \citenamefont
  {Li}(2006)}]{hou_dynamic_2006}%
  \BibitemOpen
  \bibfield  {author} {\bibinfo {author} {\bibfnamefont {T.~Y.}\ \bibnamefont
  {Hou}}\ and\ \bibinfo {author} {\bibfnamefont {R.}~\bibnamefont {Li}},\
  }\href {\doibase 10.1007/s00332-006-0800-3} {\bibfield  {journal} {\bibinfo
  {journal} {J Nonlinear Sci}\ }\textbf {\bibinfo {volume} {16}},\ \bibinfo
  {pages} {639} (\bibinfo {year} {2006})}\BibitemShut {NoStop}%
\bibitem [{\citenamefont {Luo}\ and\ \citenamefont
  {Hou}(2014)}]{luo_potentially_2014}%
  \BibitemOpen
  \bibfield  {author} {\bibinfo {author} {\bibfnamefont {G.}~\bibnamefont
  {Luo}}\ and\ \bibinfo {author} {\bibfnamefont {T.~Y.}\ \bibnamefont {Hou}},\
  }\href {\doibase 10.1073/pnas.1405238111} {\bibfield  {journal} {\bibinfo
  {journal} {Proc. Natl. Acad. Sci. U.S.A.}\ }\textbf {\bibinfo {volume}
  {111}},\ \bibinfo {pages} {12968} (\bibinfo {year} {2014})}\BibitemShut
  {NoStop}%
\bibitem [{\citenamefont {Moore}(1979)}]{moore_spontaneous_1979}%
  \BibitemOpen
  \bibfield  {author} {\bibinfo {author} {\bibfnamefont {D.~W.}\ \bibnamefont
  {Moore}},\ }\href {https://www.jstor.org/stable/79812} {\bibfield  {journal}
  {\bibinfo  {journal} {Proceedings of the Royal Society of London. Series A,
  Mathematical and Physical Sciences}\ }\textbf {\bibinfo {volume} {365}},\
  \bibinfo {pages} {105} (\bibinfo {year} {1979})}\BibitemShut {NoStop}%
\bibitem [{\citenamefont {Morf}\ \emph {et~al.}(1980)\citenamefont {Morf},
  \citenamefont {Orszag},\ and\ \citenamefont
  {Frisch}}]{morf_spontaneous_1980}%
  \BibitemOpen
  \bibfield  {author} {\bibinfo {author} {\bibfnamefont {R.~H.}\ \bibnamefont
  {Morf}}, \bibinfo {author} {\bibfnamefont {S.~A.}\ \bibnamefont {Orszag}}, \
  and\ \bibinfo {author} {\bibfnamefont {U.}~\bibnamefont {Frisch}},\ }\href
  {\doibase 10.1103/PhysRevLett.44.572} {\bibfield  {journal} {\bibinfo
  {journal} {Phys. Rev. Lett.}\ }\textbf {\bibinfo {volume} {44}},\ \bibinfo
  {pages} {572} (\bibinfo {year} {1980})}\BibitemShut {NoStop}%
\bibitem [{\citenamefont {Pelz}\ and\ \citenamefont
  {Gulak}(1997)}]{pelz_evidence_1997}%
  \BibitemOpen
  \bibfield  {author} {\bibinfo {author} {\bibfnamefont {R.~B.}\ \bibnamefont
  {Pelz}}\ and\ \bibinfo {author} {\bibfnamefont {Y.}~\bibnamefont {Gulak}},\
  }\href {\doibase 10.1103/PhysRevLett.79.4998} {\bibfield  {journal} {\bibinfo
   {journal} {Phys. Rev. Lett.}\ }\textbf {\bibinfo {volume} {79}},\ \bibinfo
  {pages} {4998} (\bibinfo {year} {1997})}\BibitemShut {NoStop}%
\bibitem [{\citenamefont {Gulak}\ and\ \citenamefont
  {Pelz}(2005)}]{gulak_high-symmetry_2005}%
  \BibitemOpen
  \bibfield  {author} {\bibinfo {author} {\bibfnamefont {Y.}~\bibnamefont
  {Gulak}}\ and\ \bibinfo {author} {\bibfnamefont {R.}~\bibnamefont {Pelz}},\
  }\href {\doibase 10.1016/J.FLUIDDYN.2004.07.004} {\bibfield  {journal}
  {\bibinfo  {journal} {Fluid Dyn. Res.}\ }\textbf {\bibinfo {volume} {36}},\
  \bibinfo {pages} {211} (\bibinfo {year} {2005})}\BibitemShut {NoStop}%
\bibitem [{\citenamefont {Chorin}(1982)}]{chorin_evolution_1982}%
  \BibitemOpen
  \bibfield  {author} {\bibinfo {author} {\bibfnamefont {A.~J.}\ \bibnamefont
  {Chorin}},\ }\href {\doibase 10.1007/BF01208714} {\bibfield  {journal}
  {\bibinfo  {journal} {Commun.Math. Phys.}\ }\textbf {\bibinfo {volume}
  {83}},\ \bibinfo {pages} {517} (\bibinfo {year} {1982})}\BibitemShut
  {NoStop}%
\bibitem [{\citenamefont {Siggia}(1985)}]{siggia_collapse_1985}%
  \BibitemOpen
  \bibfield  {author} {\bibinfo {author} {\bibfnamefont {E.~D.}\ \bibnamefont
  {Siggia}},\ }\href {\doibase 10.1063/1.865047} {\bibfield  {journal}
  {\bibinfo  {journal} {The Physics of Fluids}\ }\textbf {\bibinfo {volume}
  {28}},\ \bibinfo {pages} {794} (\bibinfo {year} {1985})}\BibitemShut
  {NoStop}%
\bibitem [{\citenamefont {Pumir}\ and\ \citenamefont
  {Siggia}(1990)}]{pumir_collapsing_1990}%
  \BibitemOpen
  \bibfield  {author} {\bibinfo {author} {\bibfnamefont {A.}~\bibnamefont
  {Pumir}}\ and\ \bibinfo {author} {\bibfnamefont {E.}~\bibnamefont {Siggia}},\
  }\href {\doibase 10.1063/1.857824} {\bibfield  {journal} {\bibinfo  {journal}
  {Physics of Fluids A: Fluid Dynamics}\ }\textbf {\bibinfo {volume} {2}},\
  \bibinfo {pages} {220} (\bibinfo {year} {1990})}\BibitemShut {NoStop}%
\bibitem [{\citenamefont {Bell}\ and\ \citenamefont
  {Marcus}(1992)}]{bell_vorticity_1992}%
  \BibitemOpen
  \bibfield  {author} {\bibinfo {author} {\bibfnamefont {J.~B.}\ \bibnamefont
  {Bell}}\ and\ \bibinfo {author} {\bibfnamefont {D.~L.}\ \bibnamefont
  {Marcus}},\ }\href {\doibase 10.1007/BF02096593} {\bibfield  {journal}
  {\bibinfo  {journal} {Commun.Math. Phys.}\ }\textbf {\bibinfo {volume}
  {147}},\ \bibinfo {pages} {371} (\bibinfo {year} {1992})}\BibitemShut
  {NoStop}%
\bibitem [{\citenamefont {Pelz}(1997)}]{pelz_locally_1997}%
  \BibitemOpen
  \bibfield  {author} {\bibinfo {author} {\bibfnamefont {R.~B.}\ \bibnamefont
  {Pelz}},\ }\href {\doibase 10.1103/PhysRevE.55.1617} {\bibfield  {journal}
  {\bibinfo  {journal} {Phys. Rev. E}\ }\textbf {\bibinfo {volume} {55}},\
  \bibinfo {pages} {1617} (\bibinfo {year} {1997})}\BibitemShut {NoStop}%
\bibitem [{\citenamefont {Grauer}\ and\ \citenamefont
  {Sideris}(1991)}]{grauer_numerical_1991}%
  \BibitemOpen
  \bibfield  {author} {\bibinfo {author} {\bibfnamefont {R.}~\bibnamefont
  {Grauer}}\ and\ \bibinfo {author} {\bibfnamefont {T.~C.}\ \bibnamefont
  {Sideris}},\ }\href {\doibase 10.1103/PhysRevLett.67.3511} {\bibfield
  {journal} {\bibinfo  {journal} {Phys. Rev. Lett.}\ }\textbf {\bibinfo
  {volume} {67}},\ \bibinfo {pages} {3511} (\bibinfo {year}
  {1991})}\BibitemShut {NoStop}%
\bibitem [{\citenamefont {Grauer}\ \emph {et~al.}(1998)\citenamefont {Grauer},
  \citenamefont {Marliani},\ and\ \citenamefont
  {Germaschewski}}]{grauer_adaptive_1998}%
  \BibitemOpen
  \bibfield  {author} {\bibinfo {author} {\bibfnamefont {R.}~\bibnamefont
  {Grauer}}, \bibinfo {author} {\bibfnamefont {C.}~\bibnamefont {Marliani}}, \
  and\ \bibinfo {author} {\bibfnamefont {K.}~\bibnamefont {Germaschewski}},\
  }\href {\doibase 10.1103/PhysRevLett.80.4177} {\bibfield  {journal} {\bibinfo
   {journal} {Phys. Rev. Lett.}\ }\textbf {\bibinfo {volume} {80}},\ \bibinfo
  {pages} {4177} (\bibinfo {year} {1998})}\BibitemShut {NoStop}%
\bibitem [{\citenamefont {Orlandi}\ and\ \citenamefont
  {Carnevale}(2007)}]{orlandi_nonlinear_2007}%
  \BibitemOpen
  \bibfield  {author} {\bibinfo {author} {\bibfnamefont {P.}~\bibnamefont
  {Orlandi}}\ and\ \bibinfo {author} {\bibfnamefont {G.~F.}\ \bibnamefont
  {Carnevale}},\ }\href {\doibase 10.1063/1.2732438} {\bibfield  {journal}
  {\bibinfo  {journal} {Physics of Fluids}\ }\textbf {\bibinfo {volume} {19}},\
  \bibinfo {pages} {057106} (\bibinfo {year} {2007})}\BibitemShut {NoStop}%
\bibitem [{\citenamefont {Kolluru}\ \emph {et~al.}(2022)\citenamefont
  {Kolluru}, \citenamefont {Sharma},\ and\ \citenamefont
  {Pandit}}]{kolluru_insights_2022}%
  \BibitemOpen
  \bibfield  {author} {\bibinfo {author} {\bibfnamefont {S.~S.~V.}\
  \bibnamefont {Kolluru}}, \bibinfo {author} {\bibfnamefont {P.}~\bibnamefont
  {Sharma}}, \ and\ \bibinfo {author} {\bibfnamefont {R.}~\bibnamefont
  {Pandit}},\ }\href {\doibase 10.1103/PhysRevE.105.065107} {\bibfield
  {journal} {\bibinfo  {journal} {Phys. Rev. E}\ }\textbf {\bibinfo {volume}
  {105}},\ \bibinfo {pages} {065107} (\bibinfo {year} {2022})}\BibitemShut
  {NoStop}%
\bibitem [{\citenamefont {Beale}\ \emph {et~al.}(1984)\citenamefont {Beale},
  \citenamefont {Kato},\ and\ \citenamefont {Majda}}]{beale_remarks_1984}%
  \BibitemOpen
  \bibfield  {author} {\bibinfo {author} {\bibfnamefont {J.~T.}\ \bibnamefont
  {Beale}}, \bibinfo {author} {\bibfnamefont {T.}~\bibnamefont {Kato}}, \ and\
  \bibinfo {author} {\bibfnamefont {A.}~\bibnamefont {Majda}},\ }\href
  {\doibase 10.1007/BF01212349} {\bibfield  {journal} {\bibinfo  {journal}
  {Commun.Math. Phys.}\ }\textbf {\bibinfo {volume} {94}},\ \bibinfo {pages}
  {61} (\bibinfo {year} {1984})}\BibitemShut {NoStop}%
\bibitem [{\citenamefont {Ponce}(1985)}]{ponce_remarks_1985}%
  \BibitemOpen
  \bibfield  {author} {\bibinfo {author} {\bibfnamefont {G.}~\bibnamefont
  {Ponce}},\ }\href {\doibase 10.1007/BF01205787} {\bibfield  {journal}
  {\bibinfo  {journal} {Communications in Mathematical Physics}\ }\textbf
  {\bibinfo {volume} {98}},\ \bibinfo {pages} {349} (\bibinfo {year}
  {1985})}\BibitemShut {NoStop}%
\bibitem [{\citenamefont {Constantin}\ \emph {et~al.}(1996)\citenamefont
  {Constantin}, \citenamefont {Fefferman},\ and\ \citenamefont
  {Majda}}]{constantin_geometric_1996}%
  \BibitemOpen
  \bibfield  {author} {\bibinfo {author} {\bibfnamefont {P.}~\bibnamefont
  {Constantin}}, \bibinfo {author} {\bibfnamefont {C.}~\bibnamefont
  {Fefferman}}, \ and\ \bibinfo {author} {\bibfnamefont {A.~J.}\ \bibnamefont
  {Majda}},\ }\href
  {https://www.osti.gov/biblio/441146-geometric-constraints-potentially-singular-solutions-euler-equations}
  {\bibfield  {journal} {\bibinfo  {journal} {Commun Part Diff Eq}\ }\textbf
  {\bibinfo {volume} {21}} (\bibinfo {year} {1996})}\BibitemShut {NoStop}%
\bibitem [{\citenamefont {Constantin}(2008)}]{constantin_singular_2008}%
  \BibitemOpen
  \bibfield  {author} {\bibinfo {author} {\bibfnamefont {P.}~\bibnamefont
  {Constantin}},\ }\href {\doibase 10.1016/j.physd.2008.01.006} {\bibfield
  {journal} {\bibinfo  {journal} {Physica D: Nonlinear Phenomena}\ }\textbf
  {\bibinfo {volume} {237}},\ \bibinfo {pages} {1926} (\bibinfo {year}
  {2008})}\BibitemShut {NoStop}%
\bibitem [{\citenamefont {Eyink}(2008)}]{eyink_dissipative_2008}%
  \BibitemOpen
  \bibfield  {author} {\bibinfo {author} {\bibfnamefont {G.~L.}\ \bibnamefont
  {Eyink}},\ }\href {\doibase 10.1016/j.physd.2008.02.005} {\bibfield
  {journal} {\bibinfo  {journal} {Physica D: Nonlinear Phenomena}\ }\bibinfo
  {series} {Euler {Equations}: 250 {Years} {On}},\ \textbf {\bibinfo {volume}
  {237}},\ \bibinfo {pages} {1956} (\bibinfo {year} {2008})}\BibitemShut
  {NoStop}%
\bibitem [{\citenamefont {Chae}(2007)}]{chae_finite-time_2007}%
  \BibitemOpen
  \bibfield  {author} {\bibinfo {author} {\bibfnamefont {D.}~\bibnamefont
  {Chae}},\ }\href {\doibase 10.1002/cpa.20138} {\bibfield  {journal} {\bibinfo
   {journal} {Communications on Pure and Applied Mathematics}\ }\textbf
  {\bibinfo {volume} {60}},\ \bibinfo {pages} {597} (\bibinfo {year}
  {2007})}\BibitemShut {NoStop}%
\bibitem [{\citenamefont {Deng}\ \emph {et~al.}(2005)\citenamefont {Deng},
  \citenamefont {Hou},\ and\ \citenamefont {Yu}}]{deng_geometric_2005}%
  \BibitemOpen
  \bibfield  {author} {\bibinfo {author} {\bibfnamefont {J.}~\bibnamefont
  {Deng}}, \bibinfo {author} {\bibfnamefont {T.~Y.}\ \bibnamefont {Hou}}, \
  and\ \bibinfo {author} {\bibfnamefont {X.}~\bibnamefont {Yu}},\ }\href
  {\doibase 10.1081/PDE-200044488} {\bibfield  {journal} {\bibinfo  {journal}
  {Communications in Partial Differential Equations}\ }\textbf {\bibinfo
  {volume} {30}},\ \bibinfo {pages} {225} (\bibinfo {year} {2005})}\BibitemShut
  {NoStop}%
\bibitem [{\citenamefont {Deng}\ \emph {et~al.}(2006)\citenamefont {Deng},
  \citenamefont {Hou},\ and\ \citenamefont {Yu}}]{deng_improved_2006}%
  \BibitemOpen
  \bibfield  {author} {\bibinfo {author} {\bibfnamefont {J.}~\bibnamefont
  {Deng}}, \bibinfo {author} {\bibfnamefont {T.~Y.}\ \bibnamefont {Hou}}, \
  and\ \bibinfo {author} {\bibfnamefont {X.}~\bibnamefont {Yu}},\ }\href
  {\doibase 10.1080/03605300500358152} {\bibfield  {journal} {\bibinfo
  {journal} {Communications in Partial Differential Equations}\ }\textbf
  {\bibinfo {volume} {31}},\ \bibinfo {pages} {293} (\bibinfo {year}
  {2006})}\BibitemShut {NoStop}%
\bibitem [{\citenamefont {Hou}\ and\ \citenamefont
  {Li}(2008)}]{hou_blowup_2008}%
  \BibitemOpen
  \bibfield  {author} {\bibinfo {author} {\bibfnamefont {T.~Y.}\ \bibnamefont
  {Hou}}\ and\ \bibinfo {author} {\bibfnamefont {R.}~\bibnamefont {Li}},\
  }\href {\doibase 10.1016/j.physd.2008.01.018} {\bibfield  {journal} {\bibinfo
   {journal} {Physica D: Nonlinear Phenomena}\ }\bibinfo {series} {Euler
  {Equations}: 250 {Years} {On}},\ \textbf {\bibinfo {volume} {237}},\ \bibinfo
  {pages} {1937} (\bibinfo {year} {2008})}\BibitemShut {NoStop}%
\bibitem [{\citenamefont {Canuto}\ \emph {et~al.}(2012)\citenamefont {Canuto},
  \citenamefont {Hussaini}, \citenamefont {Quarteroni},\ and\ \citenamefont
  {Zang}}]{canuto_2012}%
  \BibitemOpen
  \bibfield  {author} {\bibinfo {author} {\bibfnamefont {C.}~\bibnamefont
  {Canuto}}, \bibinfo {author} {\bibfnamefont {M.~Y.}\ \bibnamefont
  {Hussaini}}, \bibinfo {author} {\bibfnamefont {A.}~\bibnamefont
  {Quarteroni}}, \ and\ \bibinfo {author} {\bibfnamefont {T.~A.}\ \bibnamefont
  {Zang}, \bibfnamefont {Jr}},\ }\href@noop {} {{\selectlanguage {english}\emph
  {\bibinfo {title} {Spectral {Methods} in {Fluid} {Dynamics}}}}}\ (\bibinfo
  {publisher} {Springer Science \& Business Media},\ \bibinfo {year}
  {2012})\BibitemShut {NoStop}%
\bibitem [{\citenamefont {Leveque}(1990)}]{leveque_1990}%
  \BibitemOpen
  \bibfield  {author} {\bibinfo {author} {\bibfnamefont {R.~J.}\ \bibnamefont
  {Leveque}},\ }\href@noop {} {{\selectlanguage {english}\emph {\bibinfo
  {title} {Numerical {Methods} for {Conservation} {Laws}}}}}\ (\bibinfo
  {publisher} {Birkhäuser Verlag},\ \bibinfo {year} {1990})\BibitemShut
  {NoStop}%
\bibitem [{\citenamefont {Ray}\ \emph {et~al.}(2011)\citenamefont {Ray},
  \citenamefont {Frisch}, \citenamefont {Nazarenko},\ and\ \citenamefont
  {Matsumoto}}]{ray_resonance_2011}%
  \BibitemOpen
  \bibfield  {author} {\bibinfo {author} {\bibfnamefont {S.~S.}\ \bibnamefont
  {Ray}}, \bibinfo {author} {\bibfnamefont {U.}~\bibnamefont {Frisch}},
  \bibinfo {author} {\bibfnamefont {S.}~\bibnamefont {Nazarenko}}, \ and\
  \bibinfo {author} {\bibfnamefont {T.}~\bibnamefont {Matsumoto}},\ }\href
  {\doibase 10.1103/PhysRevE.84.016301} {\bibfield  {journal} {\bibinfo
  {journal} {Phys. Rev. E}\ }\textbf {\bibinfo {volume} {84}},\ \bibinfo
  {pages} {016301} (\bibinfo {year} {2011})}\BibitemShut {NoStop}%
\bibitem [{\citenamefont {Venkataraman}\ and\ \citenamefont
  {Ray}(2017)}]{venkataraman_onset_2017}%
  \BibitemOpen
  \bibfield  {author} {\bibinfo {author} {\bibfnamefont {D.}~\bibnamefont
  {Venkataraman}}\ and\ \bibinfo {author} {\bibfnamefont {S.~S.}\ \bibnamefont
  {Ray}},\ }\href {\doibase 10.1098/rspa.2016.0585} {\bibfield  {journal}
  {\bibinfo  {journal} {Proceedings of the Royal Society A: Mathematical,
  Physical and Engineering Sciences}\ }\textbf {\bibinfo {volume} {473}},\
  \bibinfo {pages} {20160585} (\bibinfo {year} {2017})}\BibitemShut {NoStop}%
\bibitem [{\citenamefont {Sulem}\ \emph {et~al.}(1983)\citenamefont {Sulem},
  \citenamefont {Sulem},\ and\ \citenamefont {Frisch}}]{sulem_tracing_1983}%
  \BibitemOpen
  \bibfield  {author} {\bibinfo {author} {\bibfnamefont {C.}~\bibnamefont
  {Sulem}}, \bibinfo {author} {\bibfnamefont {P.~L.}\ \bibnamefont {Sulem}}, \
  and\ \bibinfo {author} {\bibfnamefont {H.}~\bibnamefont {Frisch}},\ }\href
  {\doibase 10.1016/0021-9991(83)90045-1} {\bibfield  {journal} {\bibinfo
  {journal} {Journal of Computational Physics}\ }\textbf {\bibinfo {volume}
  {50}},\ \bibinfo {pages} {138} (\bibinfo {year} {1983})}\BibitemShut
  {NoStop}%
\bibitem [{\citenamefont {Bustamante}\ and\ \citenamefont
  {Brachet}(2012)}]{bustamante_interplay_2012}%
  \BibitemOpen
  \bibfield  {author} {\bibinfo {author} {\bibfnamefont {M.}~\bibnamefont
  {Bustamante}}\ and\ \bibinfo {author} {\bibfnamefont {M.-E.}\ \bibnamefont
  {Brachet}},\ }\href {\doibase 10.1103/PhysRevE.86.066302} {\bibfield
  {journal} {\bibinfo  {journal} {Physical review. E, Statistical, nonlinear,
  and soft matter physics}\ }\textbf {\bibinfo {volume} {86}},\ \bibinfo
  {pages} {066302} (\bibinfo {year} {2012})}\BibitemShut {NoStop}%
\bibitem [{\citenamefont {Pereira}\ \emph {et~al.}(2013)\citenamefont
  {Pereira}, \citenamefont {Nguyen~van yen}, \citenamefont {Farge},\ and\
  \citenamefont {Schneider}}]{pereira_wavelet_2013}%
  \BibitemOpen
  \bibfield  {author} {\bibinfo {author} {\bibfnamefont {R.~M.}\ \bibnamefont
  {Pereira}}, \bibinfo {author} {\bibfnamefont {R.}~\bibnamefont {Nguyen~van
  yen}}, \bibinfo {author} {\bibfnamefont {M.}~\bibnamefont {Farge}}, \ and\
  \bibinfo {author} {\bibfnamefont {K.}~\bibnamefont {Schneider}},\ }\href
  {\doibase 10.1103/PhysRevE.87.033017} {\bibfield  {journal} {\bibinfo
  {journal} {Phys. Rev. E}\ }\textbf {\bibinfo {volume} {87}},\ \bibinfo
  {pages} {033017} (\bibinfo {year} {2013})}\BibitemShut {NoStop}%
\bibitem [{\citenamefont {Farge}\ \emph {et~al.}(2017)\citenamefont {Farge},
  \citenamefont {Okamoto}, \citenamefont {Schneider},\ and\ \citenamefont
  {Yoshimatsu}}]{farge_wavelet-based_2017}%
  \BibitemOpen
  \bibfield  {author} {\bibinfo {author} {\bibfnamefont {M.}~\bibnamefont
  {Farge}}, \bibinfo {author} {\bibfnamefont {N.}~\bibnamefont {Okamoto}},
  \bibinfo {author} {\bibfnamefont {K.}~\bibnamefont {Schneider}}, \ and\
  \bibinfo {author} {\bibfnamefont {K.}~\bibnamefont {Yoshimatsu}},\ }\href
  {\doibase 10.1103/PhysRevE.96.063119} {\bibfield  {journal} {\bibinfo
  {journal} {Phys. Rev. E}\ }\textbf {\bibinfo {volume} {96}},\ \bibinfo
  {pages} {063119} (\bibinfo {year} {2017})}\BibitemShut {NoStop}%
\bibitem [{\citenamefont {Murugan}\ \emph {et~al.}(2020)\citenamefont
  {Murugan}, \citenamefont {Frisch}, \citenamefont {Nazarenko}, \citenamefont
  {Besse},\ and\ \citenamefont {Ray}}]{murugan_suppressing_2020}%
  \BibitemOpen
  \bibfield  {author} {\bibinfo {author} {\bibfnamefont {S.~D.}\ \bibnamefont
  {Murugan}}, \bibinfo {author} {\bibfnamefont {U.}~\bibnamefont {Frisch}},
  \bibinfo {author} {\bibfnamefont {S.}~\bibnamefont {Nazarenko}}, \bibinfo
  {author} {\bibfnamefont {N.}~\bibnamefont {Besse}}, \ and\ \bibinfo {author}
  {\bibfnamefont {S.~S.}\ \bibnamefont {Ray}},\ }\href {\doibase
  10.1103/PhysRevResearch.2.033202} {\bibfield  {journal} {\bibinfo  {journal}
  {Phys. Rev. Research}\ }\textbf {\bibinfo {volume} {2}},\ \bibinfo {pages}
  {033202} (\bibinfo {year} {2020})}\BibitemShut {NoStop}%
\bibitem [{\citenamefont {Pereira}\ \emph {et~al.}(2021)\citenamefont
  {Pereira}, \citenamefont {van yen}, \citenamefont {Schneider},\ and\
  \citenamefont {Farge}}]{pereira_adaptive_2021}%
  \BibitemOpen
  \bibfield  {author} {\bibinfo {author} {\bibfnamefont {R.~M.}\ \bibnamefont
  {Pereira}}, \bibinfo {author} {\bibfnamefont {N.~N.}\ \bibnamefont {van
  yen}}, \bibinfo {author} {\bibfnamefont {K.}~\bibnamefont {Schneider}}, \
  and\ \bibinfo {author} {\bibfnamefont {M.}~\bibnamefont {Farge}},\ }\href
  {http://arxiv.org/abs/2111.04863} {\bibfield  {journal} {\bibinfo  {journal}
  {arXiv:2111.04863 [physics]}\ } (\bibinfo {year} {2021})}\BibitemShut
  {NoStop}%
\bibitem [{\citenamefont {Bos}\ and\ \citenamefont
  {Bertoglio}(2006)}]{bos_dynamics_2006}%
  \BibitemOpen
  \bibfield  {author} {\bibinfo {author} {\bibfnamefont {W.~J.~T.}\
  \bibnamefont {Bos}}\ and\ \bibinfo {author} {\bibfnamefont {J.-P.}\
  \bibnamefont {Bertoglio}},\ }\href {\doibase 10.1063/1.2219766} {\bibfield
  {journal} {\bibinfo  {journal} {Physics of Fluids}\ }\textbf {\bibinfo
  {volume} {18}},\ \bibinfo {pages} {071701} (\bibinfo {year}
  {2006})}\BibitemShut {NoStop}%
\bibitem [{\citenamefont {Krstulovic}\ \emph {et~al.}(2009)\citenamefont
  {Krstulovic}, \citenamefont {Mininni}, \citenamefont {Brachet},\ and\
  \citenamefont {Pouquet}}]{krstulovic_cascades_2009}%
  \BibitemOpen
  \bibfield  {author} {\bibinfo {author} {\bibfnamefont {G.}~\bibnamefont
  {Krstulovic}}, \bibinfo {author} {\bibfnamefont {P.~D.}\ \bibnamefont
  {Mininni}}, \bibinfo {author} {\bibfnamefont {M.~E.}\ \bibnamefont
  {Brachet}}, \ and\ \bibinfo {author} {\bibfnamefont {A.}~\bibnamefont
  {Pouquet}},\ }\href {\doibase 10.1103/PhysRevE.79.056304} {\bibfield
  {journal} {\bibinfo  {journal} {Phys. Rev. E}\ }\textbf {\bibinfo {volume}
  {79}},\ \bibinfo {pages} {056304} (\bibinfo {year} {2009})}\BibitemShut
  {NoStop}%
\bibitem [{\citenamefont {Majda}\ and\ \citenamefont
  {Timofeyev}(2000)}]{majda_remarkable_2000}%
  \BibitemOpen
  \bibfield  {author} {\bibinfo {author} {\bibfnamefont {A.~J.}\ \bibnamefont
  {Majda}}\ and\ \bibinfo {author} {\bibfnamefont {I.}~\bibnamefont
  {Timofeyev}},\ }\href {\doibase 10.1073/pnas.230433997} {\bibfield  {journal}
  {\bibinfo  {journal} {Proceedings of the National Academy of Sciences}\
  }\textbf {\bibinfo {volume} {97}},\ \bibinfo {pages} {12413} (\bibinfo {year}
  {2000})},\ \Eprint
  {http://arxiv.org/abs/https://www.pnas.org/doi/pdf/10.1073/pnas.230433997}
  {https://www.pnas.org/doi/pdf/10.1073/pnas.230433997} \BibitemShut {NoStop}%
\bibitem [{\citenamefont {Clark~di Leoni}\ \emph {et~al.}(2017)\citenamefont
  {Clark~di Leoni}, \citenamefont {Mininni},\ and\ \citenamefont
  {Brachet}}]{clark_di_leoni_dynamics_2017}%
  \BibitemOpen
  \bibfield  {author} {\bibinfo {author} {\bibfnamefont {P.}~\bibnamefont
  {Clark~di Leoni}}, \bibinfo {author} {\bibfnamefont {P.}~\bibnamefont
  {Mininni}}, \ and\ \bibinfo {author} {\bibfnamefont {M.-E.}\ \bibnamefont
  {Brachet}},\ }\href {\doibase 10.1103/PhysRevFluids.3.014603} {\bibfield
  {journal} {\bibinfo  {journal} {Physical Review Fluids}\ }\textbf {\bibinfo
  {volume} {3}} (\bibinfo {year} {2017}),\
  10.1103/PhysRevFluids.3.014603}\BibitemShut {NoStop}%
\bibitem [{\citenamefont {Chorin}(1986)}]{chorin_turbulence_1986}%
  \BibitemOpen
  \bibfield  {author} {\bibinfo {author} {\bibfnamefont {A.~J.}\ \bibnamefont
  {Chorin}},\ }\href {\doibase 10.1002/cpa.3160390706} {\bibfield  {journal}
  {\bibinfo  {journal} {Communications on Pure and Applied Mathematics}\
  }\textbf {\bibinfo {volume} {39}},\ \bibinfo {pages} {S47} (\bibinfo {year}
  {1986})}\BibitemShut {NoStop}%
\bibitem [{\citenamefont {Hamlington}\ \emph {et~al.}(2008)\citenamefont
  {Hamlington}, \citenamefont {Schumacher},\ and\ \citenamefont
  {Dahm}}]{Hamlington}%
  \BibitemOpen
  \bibfield  {author} {\bibinfo {author} {\bibfnamefont {P.~E.}\ \bibnamefont
  {Hamlington}}, \bibinfo {author} {\bibfnamefont {J.}~\bibnamefont
  {Schumacher}}, \ and\ \bibinfo {author} {\bibfnamefont {W.~J.~A.}\
  \bibnamefont {Dahm}},\ }\href {\doibase 10.1103/PhysRevE.77.026303}
  {\bibfield  {journal} {\bibinfo  {journal} {Phys. Rev. E}\ }\textbf {\bibinfo
  {volume} {77}},\ \bibinfo {pages} {026303} (\bibinfo {year}
  {2008})}\BibitemShut {NoStop}%
\end{thebibliography}%
\end{document}